\documentclass[aip,jcp,reprint]{revtex4-1}

\usepackage{amsmath}
\usepackage{graphicx}
\usepackage{dcolumn}
\usepackage{bm}
\usepackage{mathrsfs}
\usepackage{mhchem}
\usepackage{subfigure}
\usepackage{color}
\usepackage{soul}

\newcommand*\oline[1]{%
   \vbox{%
     \hrule height 0.5pt
     \kern0.5ex
     \hbox{%
       \kern-0.1em
       \ifmmode#1\else\ensuremath{#1}\fi
       \kern-0.1em
     }
   }
}

\begin{document}

\title{Dynamics of Janus motors with microscopically reversible kinetics}
\author{Mu-Jie Huang}
\email{mjhuang@chem.utoronto.ca}
\affiliation{ Chemical Physics Theory Group, Department of Chemistry, University of Toronto, Toronto, Ontario M5S 3H6, Canada}

\author{Jeremy Schofield}
\email{jmschofi@chem.utoronto.ca}
\affiliation{ Chemical Physics Theory Group, Department of Chemistry, University of Toronto, Toronto, Ontario M5S 3H6, Canada}

\author{Pierre Gaspard}
\email{gaspard@ulb.ac.be}
\affiliation{ Center for Nonlinear Phenomena and Complex Systems, Universit{\'e} Libre de Bruxelles (U.L.B.), Code Postal 231, Campus Plaine, B-1050 Brussels, Belgium}

\author{Raymond Kapral}
\email{rkapral@chem.utoronto.ca}
\affiliation{ Chemical Physics Theory Group, Department of Chemistry, University of Toronto, Toronto, Ontario M5S 3H6, Canada}

\date{\today}
\begin{abstract}
Janus motors with chemically active and inactive hemispheres can operate only under nonequilibrium conditions where detailed balance is broken by fluxes of chemical species that establish a nonequilibrium state. A microscopic model for reversible reactive collisions on a Janus motor surface is constructed and shown to satisfy detailed balance. The model is used to study Janus particle reactive dynamics in systems at equilibrium where generalized chemical rate laws that include time-dependent rate coefficients with power-law behavior are shown to describe reaction rates. While maintaining reversible reactions on the Janus catalytic hemisphere, the system is then driven into a nonequilibrium steady state by fluxes of chemical species that control the chemical affinity. The statistical properties of the self-propelled Janus motor in this nonequilibrium steady state are investigated and compared with predictions of a fluctuating thermodynamics theory. The model has utility beyond the examples presented here, since it allows one to explore various aspects of nonequilibrium fluctuations in systems with self-diffusiophoretic motors from a microscopic perspective.
\end{abstract}


\maketitle

\section{Introduction}

Systems of active particles are encountered often in a number of different contexts. Molecular machines perform various tasks to assist biological functions in the cell,~\cite{alberts-cell,Jones-book}
while microorganisms swim or move autonomously in different kinds of media to seek food sources.~\cite{berg75,berg04}
Synthetic molecular machines and nano/micromotors with and without moving parts have been constructed and are able to execute directed motion.~\cite{kay:07,wangbook:13,wang:13,sanchez:14} All of these machines and motors operate out of equilibrium, experience strong thermal fluctuations and obtain energy from their environment in order to move.

An often-studied synthetic motor is a spherical Janus particle with catalytic and noncatalytic hemispheres that operates by  phoretic mechanisms.~\cite{derjaguin:47,derjaguin:74,ALP82,A89,AP91,Golestanian_etal_05,kapral:13}
For the diffusiophoretic mechanism, chemical reactions on the catalytic hemisphere interconvert reagent (fuel) and product molecules and, in the process, generate inhomogeneous concentration fields of these species in the Janus particle vicinity. The system is maintained in a nonequilibrium state by fluxes of the species at the system boundaries or in the fluid phase environment. As a result of intermolecular interactions of the reactive species with the Janus motor, the fluid exerts a force on the motor that is compensated by fluid flows in the environment that lead to motor self-propulsion. Autonomous motion is possible only if the system is driven out of equilibrium.

The mean values of properties such as the motor velocity are typically computed by adopting a continuum description where the concentration and fluid velocity fields are described by reaction-diffusion and Stokes equations, respectively. However, because of the presence of strong thermal fluctuations, stochastic models are required to describe motor motion. The underlying reactive dynamical processes on the motor surface must be microscopically reversible, and the stochastic equations of motion must account for microscopic reversibility to be consistent with thermodynamics. Langevin equations of motion that satisfy these consistency requirements have been derived and used to establish nonequilibrium fluctuation formulas for diffusiophoretic Janus motors.~\cite{GK17,GK18a}

In this paper we consider the motion of Janus motors chemically-propelled by self-diffusiophoresis from a microscopic perspective. A Janus motor is built as an aggregate of catalytic and noncatalytic beads.~\cite{debuyl:13,debuyl:18} The reactive collisions on the catalytic portion of the motor are constructed to be microscopically reversible and the reactive kinetics satisfies detailed balance. The Janus particle reaction kinetics is studied both in systems at equilibrium where the Janus particle is chemically active but propulsion is not possible, as well as under nonequilibrium conditions where it is self-propelled.

The paper is structured as follows: The model for a Janus motor and reversible reactive collision dynamics on the motor catalytic surface are described in Sec.~\ref{sec:micro_model}. Section~\ref{sec:det-bal} demonstrates that the reactive dynamics satisfies the condition of microscopic reversibility and the kinetics obeys detailed balance. Simulations of the dynamics of a Janus particle in systems at equilibrium are presented in Sec.~\ref{sec:equib} where it is shown that the equilibrium reactive species number fluctuations are binomially distributed, and that chemical relaxation obeys a generalized rate law with time-dependent reaction rate coefficients. Nonequilibrium dynamics is the subject of Sec.~\ref{sec:nonequib}. The system is driven out of equilibrium by the control of concentrations of chemical species at a distant boundary. In this section the influence of an externally applied force to the motor on the reaction rate is also considered. In Sec.~\ref{sec:bulk_reaction}, in addition to reactions on the motor surface, an out-of-equilibrium fluid phase reaction is implemented to break detailed balance in the bulk phase instead of the distant boundary, while retaining the microscopically reversible reactive dynamics on the motor surface. Janus motor self-propulsion is now possible and its characteristics are studied and compared with continuum theory. The conclusions of the work are given in Sec.~\ref{sec:conc}.

\section{Janus motor system and catalytic reactions}\label{sec:micro_model}

We consider a single Janus motor immersed in a fluid of inert ($S$) and reactive ($A$ and $B$) particles.  The motor interacts directly with the fluid particles while interactions among fluid particles are taken into account by multiparticle collision dynamics.~\cite{Malevanets_Kapral_99,*Malevanets_Kapral_00}

\begin{figure}[htbp]
\centering
\resizebox{0.9\columnwidth}{!}{%
\includegraphics{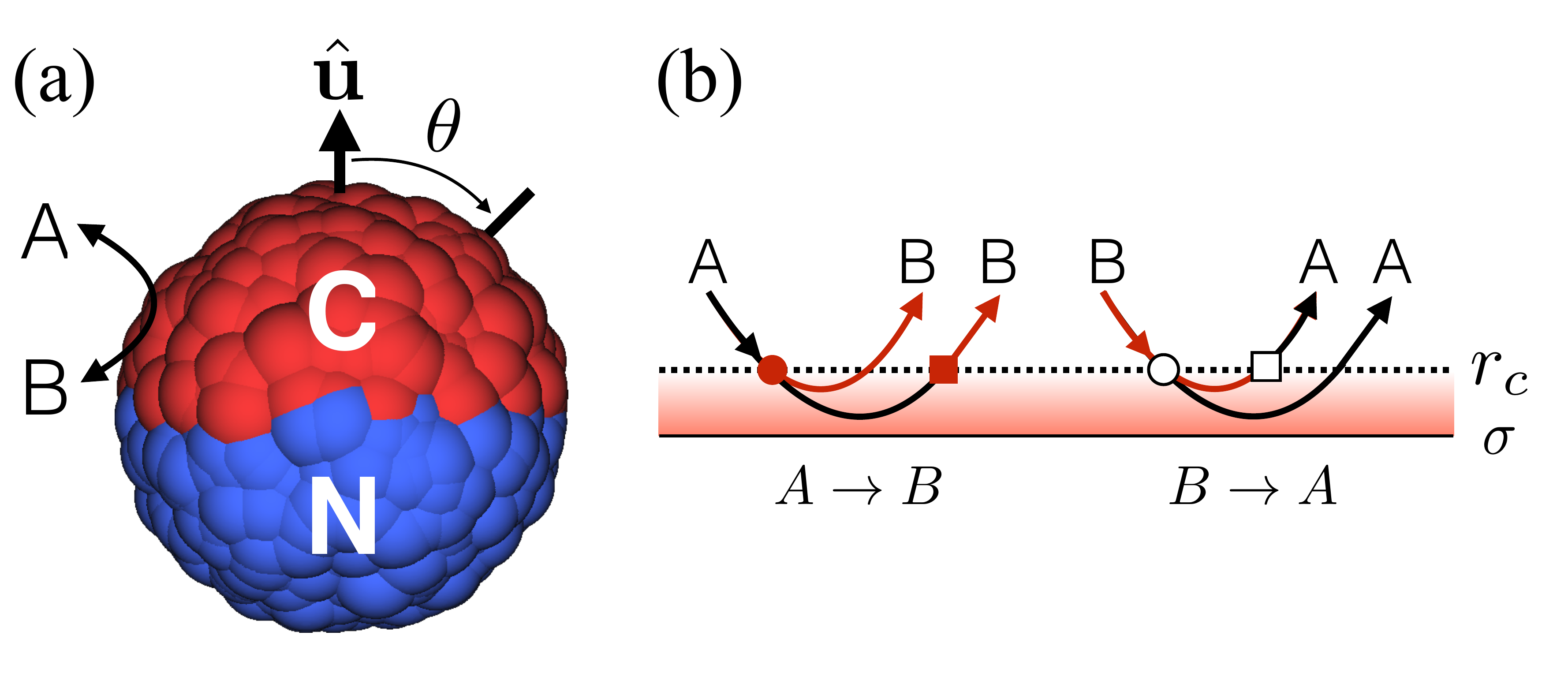}}
\caption{\label{fig:Janus}(a) A Janus motor comprising beads with radius $\sigma$ connected by springs (not shown) has catalytic ($C$, red) and noncatalytic ($N$, blue) hemispheres. The motor axis is defined by the unit vector, $\hat{\mathbf{u}}$, in the direction from the $N$ to the $C$ hemispheres, and $\theta$ is the polar angle. Reversible catalytic reactions occur when particles $A$ or $B$ encounter the motor $C$ beads. (b) Trajectories for the forward ($A \to B$) and reverse ($B\to A$) reactive collisions. In this diagram, a fuel particle $A$ (product $B$) follows a black (red) trajectory, where reactions take place upon entering (circles) into or leaving (squares) from the motor surface. The solid and hollow symbols signify the forward ($A\to B$) and reverse ($B\to A$) reactions, respectively.}
\end{figure}

The Janus motor is constructed as a roughly spherical object composed of $N_b=N_C +N_N$ of $N_C$ catalytic ($C$) and $N_N$ noncatalytic ($N$) beads with mass $m$ that differ in their interactions with the solvent particles and in their chemical activity~\cite{debuyl:13} (see Fig.~\ref{fig:Janus} (a)).

Janus motor beads $j$ and $k$ at positions $\mathbf{r}_{bj}$ and $\mathbf{r}_{bk}$ interact via a harmonic potential $U_J(r_{jk}) = \frac{1}{2} k_s (r_{jk}-r_{jk}^0)^2$, with $r_{jk}=|\mathbf{r}_{bj} - \mathbf{r}_{bk}|$, if their equilibrium distance $r_{jk}^0 < 2\:\sigma$, where $k_s$ is a stiff spring constant that ensures fluctuations of the positions of the beads are small so that the motor retains its spherical shape during the evolution of the system. The isolated Janus motor has a potential energy
\begin{equation}
U_J({\bf r}_b) = \sum_{i=1}^{N_b} \sum_{j<i} U_J(r_{ij}),
\label{janusEnergy}
\end{equation}
where ${\bf r}_b=({\bf r}_{b1},{\bf r}_{b2},\dots,{\bf r}_{bN_b})$.

The surrounding fluid consists of $N_R$ reactive $A$ and $B$ species with coordinates ${\bf r}_R =({\bf r}_{R1},{\bf r}_{R2}, \dots,{\bf r}_{R N_R})$, as well as $N_S$ chemically inert $S$ species with coordinates ${\bf r}_S =({\bf r}_{S1},{\bf r}_{S2}, \dots,{\bf r}_{S N_S})$. In this notation, the index of the particle follows the symbol $b$, $R$ or $S$ specifying the type of the particle. Collectively the coordinates of the fluid particles are ${\bf r}_{{\rm f}}=({\bf r}_R,{\bf r}_S)$ and we let ${\bf r}=({\bf r}_b, {\bf r}_{{\rm f}})$ denote all of the coordinates. The $N_R$ reactive particles have species labels $\boldsymbol{\alpha}=(\alpha_1, \alpha_2, \dots, \alpha_{N_R})$ where $\alpha_i \in \{A,B\}$. Since the inert species are all of type $S$ we do not include them in the set $\boldsymbol{\alpha}$. Letting the interaction energy between a motor bead $j$ and a solvent particle $i$ of type $\alpha = A, B, S$ be $U_{\alpha}(|{\bf r}_{{\rm f}i} -{\bf r}_{bj}|)$, the total potential energy of the fluid particles is
\begin{equation}\label{eq:pot}
U_{{\rm f}}({\bf r}, \boldsymbol{\alpha})=\sum_{i=1}^{N_R}U_{R}({\bf r}_{Ri},\alpha_i, {\bf r}_b)+\sum_{i=1}^{N_S} U_{S}({\bf r}_{Si},{\bf r}_b),
\end{equation}
with
\begin{eqnarray}\label{eq:Rpot}
U_{R}({\bf r}_{Ri},\alpha_i, {\bf r}_b)&=& \sum_{j=1}^{N_b} U_{\alpha_i}(|{\bf r}_{Ri} -{\bf r}_{bj}|),\nonumber \\
U_{S}({\bf r}_{Si}, {\bf r}_b)&=& \sum_{j=1}^{N_b} U_{S}(|{\bf r}_{Si} -{\bf r}_{bj}|).
\end{eqnarray}
In the applications discussed below $U_{\alpha}(r)$ is taken to be a repulsive Lennard-Jones potential with interaction strength $\epsilon_{\alpha}$, $U_{\alpha}(r) =  4\epsilon_{\alpha} [(\sigma/r)^{12} - (\sigma/r)^6 + 0.25]\Theta(r_c-r)$, where $\Theta(r_c-r)$ is a Heaviside function with $r_c = 2^{1/6}\sigma$. (For simplicity, here we suppose that the interaction energy does not depend on the type of motor bead. Extension to the general case is straightforward.)

The potential energy of the entire system is $U_{\rm T}({\bf r}, \boldsymbol{\alpha})= U_0 + U({\bf r},\boldsymbol{\alpha})$, where $U({\bf r},\boldsymbol{\alpha})=U_J({\bf r}_J)+ U_{{\rm f}}({\bf r}, \boldsymbol{\alpha})$, and $U_0 =\sum_{i=1}^{N_R} u^0_{\alpha_i}$ accounts for bare internal energies $u^0_{\alpha_i}$ of the reactive chemical species. The total energy is $E=K_J +K_{{\rm f}} +U_{\rm T}({\bf r}, \boldsymbol{\alpha})$, where $K_J$ and $K_{{\rm f}}$ are the kinetic energies of the motor beads and fluid particles. Notice that there are no interactions among the solvent and reactive particles.  These interaction effects are taken into account by using the multiparticle collision method.~\cite{Kapral_08} The simulation method and parameters are described in detail in Appendix~\ref{app:sim}.

\subsection*{Motor-catalyzed reactions}
Interactions of the $A$ and $B$ species with the catalytic beads may lead to the reversible chemical reaction,
\begin{equation}
C+A \underset{k_-}{\stackrel{k_+}{\rightleftharpoons}} C+B,
\label{janusReaction}
\end{equation}
where $k_{\pm}$ denote the forward and reverse reaction rate constants. A full description of reactive dynamics at the catalytic portion of the Janus motor surface requires a microscopic definition of chemical species and specification of the bond-making and bond-breaking events that constitute the chemical transformations from reactants to products. For example, a common reaction mechanism involves species interconversion dynamics governed by a double-well potential function for a reaction coordinate. The potential wells can be used to define the metastable chemical species. In the bulk phase, outside of the interaction range with the Janus motor, the barrier separating reactants from products is assumed to be very high so that chemical reactions will occur with extremely low probability. Interactions with the catalytic face of the motor cause the barrier height to be reduced thus facilitating the reactive events.

Instead of a full dynamical description, we suppose that in the bulk of the solution the constant bare potential energy functions, $u^0_{\alpha}$, ($\alpha=A,B$) are associated with the $A$ and $B$ species and characterize their internal states. Instead of describing the reactions by deterministic motion in the potential energy surface of the reactive system, we encode the likelihood of chemical transformations in probabilities $p_\pm$ for forward $A \to B$ and reverse $B \to A$ reactive events. Since the fluid species interact with the surface beads of the Janus particle through short-range intermolecular potentials, we may define a reaction surface ${\mathcal S}_R$, which depends on the Janus particle configuration ${\bf r}_b$, and outside of which interactions with the Janus catalytic beads vanish. The region interior to the reaction surface is the reaction zone. Chemical transformations between the $A$ and $B$ species may take place when these species cross the reaction surface. In the simulations the reactions occur infinitesimally outside of the reaction surface where the forces derived from the interaction potential are zero. This choice avoids difficulties in the molecular dynamics associated with sudden changes in the potential functions.

The coarse-grain reactive events take place as follows (see Fig.~\ref{fig:Janus} (b)): Reactions with the catalytic beads may occur whenever an $A$ or $B$ particle reaches a point infinitesimally outside of ${\mathcal S}_R$ at $r=r_c$.  More specifically, in our coarse-grain model a forward reaction, $A \to B$, may occur with equal probability $p_+/2$ when an $A$ particle enters or leaves from the motor reaction zone. If the forward reaction occurs as the $A$ particle enters the reaction zone (red solid circle), it will propagate as a product $B$ particle and eventually leave this zone. Similarly, if the forward reaction occurs as the $A$ particle leaves the reactive zone (red solid square), it will have propagated as an $A$ particle during its interactions with the motor catalytic beads. Since $A$ and $B$ particles have different interaction potentials with the motor catalytic beads these two reactive trajectories differ. Reactions take place only as these chemical species enter or leave the reaction zone and no additional reactive events are allowed to take place within the zone. Similarly, the reverse reaction
$B \to A$ may occur with equal probability $p_-/2$ when a $B$ particle enters (black hollow circle) or leaves (black hollow square) the zone. Moreover, we assume no change of velocities upon reaction.

\section{Reversible dynamics and detailed balance}\label{sec:det-bal}

We let $\mathbf{x}=({\bf v}, {\bf r})=(\mathbf{x}_b,\mathbf{x}_{{\rm f}})$ be the phase point of the entire system, where $\mathbf{x}_b=({\bf v}_b, {\bf r}_b)$ and $\mathbf{x}_{{\rm f}}=({\bf v}_{{\rm f}}, {\bf r}_{{\rm f}})$ with ${\bf v}_b$ and ${\bf v}_{{\rm f}}$ the set of velocities of the Janus motor beads and fluid particles, respectively. The phase space probability density is denoted by $P(\mathbf{x},\boldsymbol{\alpha},t)$ and its evolution is given by the equation of motion,
\begin{equation}\label{eq:pevol}
\frac{\partial}{\partial t} P(\mathbf{x},\boldsymbol{\alpha},t)= {\mathcal L}P(\mathbf{x},\boldsymbol{\alpha},t),
\end{equation}
where ${\mathcal L}={\mathcal L}_D + {\mathcal C} +{\mathcal L}_R$ is the sum of deterministic, multiparticle collision and reactive evolution operators. The Liouvillian ${\mathcal L}_D=-{\bf v} \cdot \mathbf{\nabla}_{{\bf r}} -({\bf F}/m) \cdot \mathbf{\nabla}_{{\bf v}}$ for deterministic evolution involves forces derived from the full potential $U({\bf r}, \boldsymbol{\alpha})$, while ${\mathcal C}$, the evolution operator for multiparticle collisions, is defined elsewhere~\cite{Kapral_08} and its explicit form will not be required here. To write the reactive Liouville operator, ${\mathcal L}_R$, corresponding to the reactive dynamics discussed above, we first let ${\bf r}_{iJ}$ and ${\bf v}_{iJ}$ denote the position and velocity of particle $i$ relative to the position ${\bf r}_J$ of the center of mass of the Janus motor. The magnitude of the vector ${\bf r}_{iJ}$ at a point infinitesimally outside the reaction surface will be denoted by $R^+(\hat{{\bf r}}_{iJ},{\bf r}_b)$ since its value depends on its location on the surface and the configuration of the Janus beads. The normal to the reaction surface at this point is denoted by $\hat{{\bf n}}(\hat{{\bf r}}_{iJ},{\bf r}_b)$. (We omit the arguments of these functions in the following.) The reactive Liouville operator may now be written as
\begin{eqnarray}
&&{\mathcal L}_R= \sum_{i=1}^{N_R} \sum_s |{\bf v}_{iJ} \cdot \hat{{\bf n}}| \Theta(s{\bf v}_{iJ} \cdot \hat{{\bf n}}) \delta(r_{iJ} - R^+) \\
&& \quad \times \frac{1}{2}[\delta_{\alpha_i A}(p_- {\mathcal E}_i^{A \to B} -p_+) + \delta_{\alpha_i B}(p_+ {\mathcal E}_i^{B \to A} -p_-)],\nonumber
\end{eqnarray}
where the index $s$ takes the values $s=\pm$ for entering or leaving the reaction zone, and the operator ${\mathcal E}_i^{\alpha \to \alpha^\prime}$ changes the species index of particle $i$ from $\alpha$ to $\alpha'$. This dynamics conserves mass, momentum and energy and we now show that the reactive dynamics satisfies detailed balance.

{\em Detailed balance}: Without loss of generality, we consider a single particle $i$ of type $\alpha$ at time $t$ that is about to cross the reactive boundary ${\mathcal S}_R$ at a point on the surface that lies at $r_{iJ}=R^+$ from the Janus particle center.  We compute the contribution to the reactive flux of species $A$ for this particle,
The trajectories contributing to this flux were discussed in Sec.~\ref{sec:micro_model}. Particle $i$ with species label $A$ converts to $B$ with probability $p_+/2$ as it enters the reaction zone. There is a corresponding trajectory, obtained by time reversal from this trajectory, that converts $B$ to $A$ with probability $p_-/2$ when it leaves the reaction zone at the reaction boundary. Similarly, particle $i$ with species label $B$ converts to $A$ with probability $p_-/2$ as it enters the reaction zone. There is a corresponding trajectory obtained by time reversal from this trajectory that converts $A$ to $B$ with probability $p_+/2$ when it leaves the reaction zone at the reaction boundary. The reactive flux may be written as
\begin{eqnarray}
&&{\mathcal R}_i^A(\mathbf{x},\pmb{\alpha},t)d\mathbf{x}= \sum_s  | \mathbf{v}_{iJ} \cdot \hat{\mathbf{n}} |
\Theta(s \mathbf{v}_{iJ} \cdot \hat{\mathbf{n}}) \delta (r_{iJ} - R^+) \nonumber \\
&& \times \frac{1}{2} \Big(p_- P(\mathbf{x},\pmb{\alpha},t|B,R^+) - p_+ P(\mathbf{x},\pmb{\alpha},t|A,R^+)  \Big)d\mathbf{x}.
\end{eqnarray}
Here $P(\mathbf{x},\pmb{\alpha},t|\alpha_i=\alpha,r_{iJ}=R^+)\equiv P(\mathbf{x},\pmb{\alpha},t|\alpha,R^+)$ is the probability density at $(\mathbf{x},\pmb{\alpha})$ at time $t$ given that particle $i$ lies at the point infinitesimally outside the reaction boundary and is species $\alpha$.

At equilibrium this expression yields the detailed balance condition,
\begin{equation}\label{eq:det-bal1}
p_+ P_{{\rm eq}}(\mathbf{x},\pmb{\alpha}|A,R^+)= p_- P_{{\rm eq}}(\mathbf{x},\pmb{\alpha}|B,R^+).
\end{equation}
This equation may be integrated over all phase space coordinates and summed over all species labels except for the position of particle $i$ and its species label. Denoting the reduced distributions that result from this integration by $P_{{\rm eq}}(\alpha,R^+)$ we obtain
\begin{equation}\label{eq:det-bal}
\frac{P_{{\rm eq}}(B,R^+)}{P_{{\rm eq}}(A,R^+)}=\frac{p_+}{p_-}=\frac{k^0_+}{k^0_-},
\end{equation}
where the last equality uses the fact that the intrinsic rate constants, $k_{\pm}^0$, are proportional to the reaction probabilities, $k_{\pm}^0 = p_{\pm} \nu_{\rm col}$, with $\nu_{\rm col}$ the collision frequency.

Under this reversible coarse-grain reactive dynamics the system will evolve to an equilibrium state with reactive solute concentrations $c^{\rm eq}_{A}$ and $c^{\rm eq}_{B}$ determined by the choice of reaction probabilities. The forces that enter the equations of motion are derived from the potential function $U({\bf r}, \boldsymbol{\alpha})$ and do not depend on the constant bare energies; the information about their values is encoded in the reaction probabilities since their values determine the equilibrium concentrations.

The equilibrium ratio $P_{{\rm eq}}(B,R^+)/P_{{\rm eq}}(A,R^+)$ can be computed as follows:
The equilibrium canonical probability density factors into Boltzmann kinetic and configurational parts. The configurational probability density takes the form,
\begin{equation}\label{eq:Peq-expression}
P_{{\rm eq}}({\bf r},\boldsymbol{\alpha})= e^{-\beta U_{\rm T}({\bf r}, \boldsymbol{\alpha})}/\sum_{\boldsymbol{\alpha}} \int  d {\bf r} \; e^{-\beta U_{\rm T}({\bf r}, \boldsymbol{\alpha})},
\end{equation}
where $\beta=(k_{\rm B}T)^{-1}$ is the inverse temperature.
To compute the left side of Eq.~(\ref{eq:det-bal}) we consider the probability density of a particle $i$ of species $\alpha'$ at a position ${\bf r}'={\bf r}_{iJ}$:
\begin{eqnarray}\label{eq:reduced-den}
P_{{\rm eq}}(\alpha',\bm{r}')&=& \sum_{\boldsymbol{\alpha}} \int  d {\bf r} \; \delta(\bm{r}'-{\bf r}_{iJ})\delta_{\alpha_i, \alpha'} P_{{\rm eq}}({\bf r},\boldsymbol{\alpha})\nonumber \\
&\equiv& e^{-\beta (u^0_{\alpha'}+u_{\alpha'}(\bm{r}'))}\Big/\sum_{\alpha'=A}^B Z_{\alpha'}.
\end{eqnarray}
The second equality defines the potential of mean force, $u_{\alpha'}(\bm{r}')$, and we have introduced the quantity $Z_\alpha=e^{-\beta u^0_{\alpha}}\int d{\bm{r}'}\:e^{-\beta u_{\alpha}(\bm{r}')}$ in writing the equation. For values of $\bm{r}'=\hat{\bm{r}}' R^+$ outside of the range of the potential $u_{\alpha'}(\hat{\bm{r}}'R^+)=0$ and, using Eq.~(\ref{eq:reduced-den}), we have
\begin{equation}
\frac{P_{{\rm eq}}(B,R^+)}{P_{{\rm eq}}(A,R^+)}=e^{-\beta \Delta u^0_{BA}},
\end{equation}
with $\Delta u^0_{BA}=u^0_{B} -u^0_{A}$. Comparison with Eq.~~(\ref{eq:det-bal}) gives $p_+/p_-=e^{-\beta \Delta u^0_{BA}}$, which shows how the reaction probabilities encode information about the bare potentials that are related to the equilibrium concentrations.

The probability of a reactive particle to be species $\alpha$ can be obtained by integration of Eq.~(\ref{eq:reduced-den}) over ${\bf r}'$ to give
\begin{equation}\label{eq:Peq}
P_{{\rm eq}}(\alpha)= Z_{\alpha}/(Z_A+Z_B),
\end{equation}
and the average number of particles of species $\alpha$ is $\langle N_{\alpha} \rangle=N_\alpha^{\rm eq}= N_R Z_{\alpha}/(Z_A+Z_B)$. We can write $Z_{\alpha} = e^{-\beta u^0_{\alpha}} V_{\alpha}= e^{-\beta u^0_{\alpha}} \gamma_{\alpha}^{-1} V$ where $V_{\alpha}$ can be interpreted as the free volume available to solvent particles of type $\alpha$, and $\gamma_{\alpha}  = V/V_{\alpha}$ is the activity coefficient of species $\alpha$.
From the definition of the activity coefficient, we find that
\begin{eqnarray}
\gamma_{\alpha}^{-1} = 1 + \frac{1}{V} \int d\bm{r}'\; \left( e^{-\beta u_{\alpha}(\bm{r}')} -1\right).
\label{activityCoefficient}
\end{eqnarray}

For short-ranged potentials, the integral in Eq.~(\ref{activityCoefficient}) is small relative to the total volume of the system and the activity coefficients are close to unity.

Since the bulk equilibrium concentration of species $\alpha$ is
$c_\alpha^{{\rm eq}} = N_{\alpha}^{{\rm eq}} /V$, using these results we have
\begin{equation}
\frac{\gamma_B c_B^{{\rm eq}}}{\gamma_A c_A^{{\rm eq}}}=\frac{a_B^{{\rm eq}}}{a_A^{{\rm eq}}}= e^{-\beta \Delta u^0_{BA}},
\label{eq:eq_ratio}
\end{equation}
where $a_\alpha^{{\rm eq}}$ is the activity of species $\alpha$. The equilibrium constant is defined by $K_{{\rm eq}} = a_B^{{\rm eq}}/a_A^{{\rm eq}}$. From these results the Guldberg-Waage form of detailed balance, $k^0_+/k^0_-= K_{{\rm eq}}$, is obtained.

\section{Janus particles in systems at equilibrium}\label{sec:equib}

\subsection{Equilibrium species number fluctuations}

\begin{figure}[htbp]
\centering
\resizebox{0.9\columnwidth}{!}{%
\includegraphics{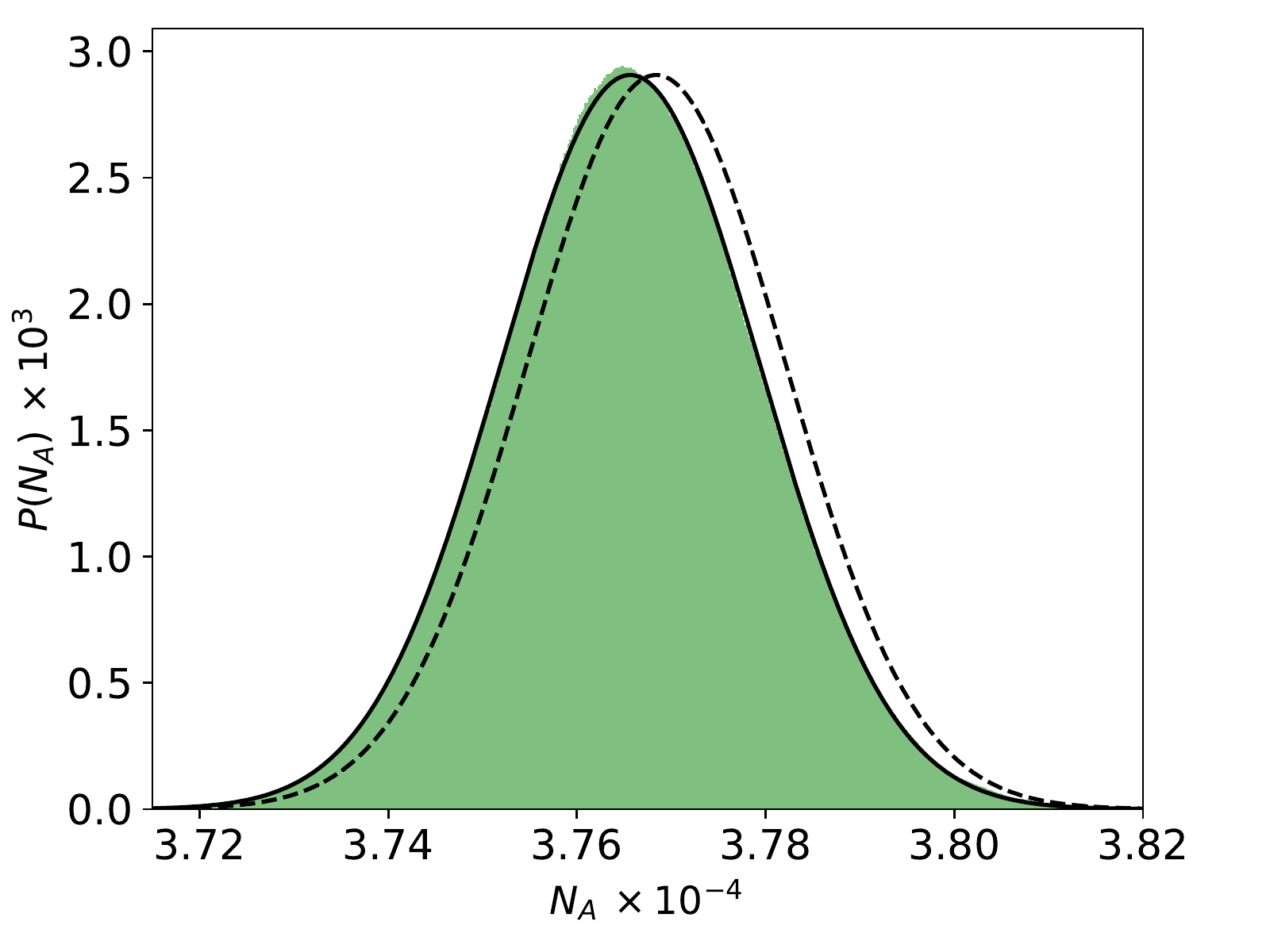}}
\caption{\label{fig:his_NA}Comparison between the histogram of the total number of $A$ particles from simulations (green solid area) and the binomial distributions with $p_A = 0.49977$ (solid curve) and $p_A = 0.5$ (dashed curve). }
\end{figure}

We consider a Janus motor where chemical reactions occur on the catalytic face with probabilities $p_{\pm} = 0.5$ in a system at equilibrium containing $N_R=N_A+N_B$ reactive solute species. The interaction strengths of the repulsive interactions between motor beads and fluid particles as described in Sec.~\ref{sec:micro_model} are $\epsilon_A = 1$, $\epsilon_B = 0.5$ and $\epsilon_S = 0.5$. We can compute the probability, $P_{{\rm eq}}(N_A)$, that there are $N_A$ particles of species $A$ in the system. Starting from an initial number of $N_A$ and $N_B$ particles ($N_A+N_B=N_R$), the system was evolved in time under the microscopic dynamics until an equilibrium state was reached. The distribution $P_{{\rm eq}}(N_A)$ was determined from a histogram of $N_A$ values and is shown in Fig.~\ref{fig:his_NA}. The function is accurately described by a binomial probability distribution,
\begin{equation} \label{eq:eqdensity-NA}
P_{{\rm eq}}(N_A)= {N_R \choose N_A} p_A^{N_A} (1-p_A)^{N_R-N_A},
\end{equation}
with mean number $N_A^{{\rm eq}}= p_A N_R$ where $p_A \approx 0.49977$, as shown in the figure.

The fact that the binomial distribution provides a highly accurate description of $P_{{\rm eq}}(N_A)$ can be understood from the following considerations. The probability density of finding a species label configuration $\boldsymbol{\alpha}$ may be obtained by integration of the equilibrium distribution (\ref{eq:Peq-expression}) over all system coordinates, $P_{{\rm eq}}(\boldsymbol{\alpha})=\int d{\bf r}\; P_{{\rm eq}}({\bf r},\boldsymbol{\alpha})$, which may be written as,
\begin{equation}\label{eq:prob-den}
P_{{\rm eq}}(\boldsymbol{\alpha})=\frac{\int d {\bf r}_b P_J({\bf r}_b) \prod_{i=1}^{N_R} Z_{\alpha_i}({\bf r}_b)
}{\sum_{\boldsymbol{\alpha}} \int d {\bf r}_b P_J({\bf r}_b) \prod_{i=1}^{N_R} Z_{\alpha_i}({\bf r}_b)  },
\end{equation}
where $P_J({\bf r}_b)$ is the effective probability density of Janus bead coordinates obtained by integrating over all solvent positions, and $Z_{\alpha_i}({\bf r}_b)= \int d {\bf r}_{Ri} e^{-\beta U_{R}({\bf r}_{Ri},\alpha_i,{\bf r}_b)}$. The dependence of the $Z_{\alpha_i}({\bf r}_b)$ factors on the Janus bead coordinates prevents this distribution from being binomial. However, if the fluctuations of the Janus particle beads are small we may suppose that their positions relative to the Janus center of mass are fixed at ${\bf r}_b^0$. Furthermore, if a reactive solute molecule interacts with only one bead (as is the case for our simulation parameters), the $Z_{\alpha_i}$ functions are independent of coordinates and we obtain,
\begin{equation}\label{eq:prob-den-approx}
P_{{\rm eq}}(\boldsymbol{\alpha})=\frac{ \prod_{i=1}^{N_R} Z_{\alpha_i}
}{\sum_{\boldsymbol{\alpha}}  \prod_{i=1}^{N_R} Z_{\alpha_i}  },
\end{equation}
and from this expression one can deduce that $P_{{\rm eq}}(N_A)$ has the binomial form given in Eq.~(\ref{eq:eqdensity-NA}). Furthermore, the $Z_{\alpha_i}$ are equal to the corresponding quantities defined below Eq.~(\ref{eq:reduced-den}) in Sec.~\ref{sec:det-bal} when the same approximations to obtain the binomial form are used to evaluate them.

From Eq.~(\ref{eq:Peq}) and the expression for $N_A^{{\rm eq}}$ below it, we have the general expression $p_A=Z_{A}/(Z_A+Z_B)=\big(1+(\gamma_A / \gamma_B) e^{-\beta \Delta u^0_{BA}}\big)^{-1}$. Since $p_\pm=0.5$ in our simulations we have $\Delta u^0_{BA}=0$ and $p_A$ takes the simpler form $p_A=(1+\gamma_A / \gamma_B )^{-1}$. The activity coefficients $\gamma_{A,B}$ can be estimated using Eq.~(\ref{activityCoefficient}).
For the Janus particle and system sizes considered in the later sections of the paper the activity coefficients can be taken to be unity. However, for smaller system sizes such as $L = 20$ in Fig.~\ref{fig:his_NA}, there are small deviations that, nevertheless, can be detected in the figure. When $L = 20$ the ratio of the activity coefficients is found to be $\gamma_A/\gamma_B \simeq 1.0009$ which yields $p_A \approx 0.49977$. One can see that this value provides a noticeably better fit than the dashed curve using $p_A = 0.5$ for unit activity coefficients.

\subsection{Reactive dynamics in systems at equilibrium}

The microscopic evolution equation for the deviation in the number of $A$ or $B$ particles in the system from their equilibrium values, $\delta N_A(t) = N_A(t) - N_A^{\rm eq}=-\delta N_B(t)$, is given by
\begin{equation}\label{eq:var-evol}
\frac{d}{d t} \delta N_A(t)= {\mathcal L}^\dagger \delta N_A(t),
\end{equation}
where ${\mathcal L}^\dagger$ is the adjoint of ${\mathcal L}$ defined in Eq.~(\ref{eq:pevol}). This equation can be cast in the form of a generalized Langevin equation using projection operator methods~\cite{zwanzigbook:01,kapral:81},
\begin{equation}\label{eq:var-evol-lang}
\frac{d}{d t} \delta N_A(t)= - \int_{0}^{t} d\tau \; \frac{\phi_k(\tau )}{V} \delta N_A(t-\tau) +f_R(t),
\end{equation}
where $f_R(t)$ is a random reaction rate with zero mean and fluctuation-dissipation relation,
\begin{equation}
\langle f_R(t)\rangle=0, \quad \frac{\phi_k(t)}{V}=\langle f_R(t) f_R(0)\rangle /\langle (\delta N_A(0))^2\rangle,
\end{equation}
where the angular brackets denote an average over $\rho_{{\rm eq}}({\bf x}, \boldsymbol{\alpha})$, the equilibrium phase space density. It has the additional property that $\langle f_R(t) \delta N_A(0)\rangle=0$.

The nonequilibrium phase space density for a system linearly displaced from chemical equilibrium is,
\begin{equation}
\rho({\bf x},\boldsymbol{\alpha}) = \rho_{{\rm eq}}({\bf x},\boldsymbol{\alpha})(1 -\delta N_A A_{{\rm rxn}} ),
\end{equation}
where $A_{{\rm rxn}}$ is the dimensionless chemical affinity. The average of Eq.~(\ref{eq:var-evol-lang}) over this nonequilibrium density yields
\begin{equation}\label{eq:rate-law-mem}
\frac{d}{dt} \langle \delta {N_A(t)}\rangle_{{\rm n}} =- \int_{0}^{t} d\tau \; \frac{\phi_k(\tau )}{V}  \langle \delta {N_A(t-\tau)}\rangle_{{\rm n}},
\end{equation}
where the angular brackets with subscript ${\rm n}$ denote the nonequilibrium average.

Alternatively, we may construct an evolution equation for the autocorrelation function of the equilibrium fluctuations of the particle number, $C_{AA}(t) = \langle \delta N_A(t) \delta N_A(0)\rangle$, by multiplying Eq.~(\ref{eq:var-evol-lang}) by
$\delta N_A(0)$ and averaging over the equilibrium density to obtain,
\begin{equation}\label{eq:rate-law-mem-C}
\frac{d}{dt}C_{AA}(t) =- \int_{0}^{t} d\tau \; \frac{\phi_k(\tau )}{V}  C_{AA}(t-\tau).
\end{equation}
From these results, in accord with the Onsager regression hypothesis~\cite{Onsager:31a,*Onsager:31b}, the regression of the microscopic fluctuations of $\delta N_A(t)$ at equilibrium should obey the same macroscopic law as the relaxation of $\langle \delta {N_A(t)}\rangle_{{\rm n}}$.

The memory kernel $\phi_k(\tau)$ evolves on a microscopic time scale $t_{{\rm mic}}$ that is much shorter than that of the chemical relaxation time $t_{{\rm chem}}$ of $C_{AA}(t)$. In such a circumstance, where $t_{{\rm mic}} \ll t_{{\rm chem}}$, the generalized rate law takes the form,
\begin{equation}\label{eq:tdep-coef}
\frac{d}{dt}C_{AA}(t) \approx -  \frac{k(t)}{V}C_{AA}(t),
\end{equation}
where the time-dependent rate coefficient is defined by
\begin{equation}
k(t)=\int_0^t d\tau \; \phi_k(\tau).
\end{equation}
The factor $1/V$ in these equations accounts for the concentration of the single Janus particle in the volume $V$.

\begin{figure}[htbp]
\centering
\resizebox{0.9\columnwidth}{!}{%
\includegraphics{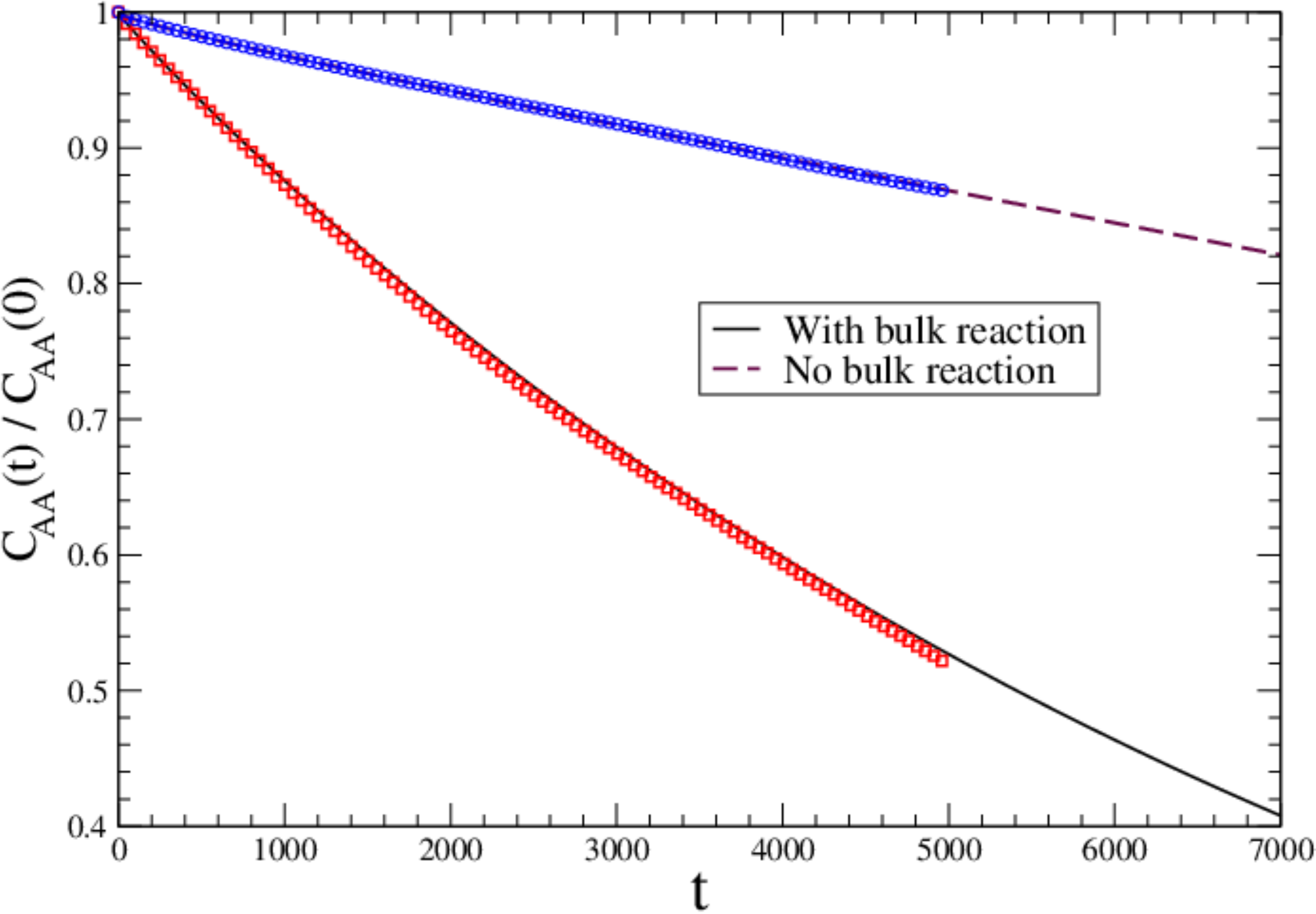}}
\caption{\label{fig:chi_t} (upper curve) Plot of the normalized autocorrelation function $C_{AA}(t)/C_{AA}(0)$ versus dimensionless simulation time obtained from the microscopic simulation of the dynamics with motor reaction probabilities $p_+=p_-=0.5$ (blue circles). These results are compared to those using the numerical Laplace inversion of Eq.~(\ref{Na_z-nb}) (dashed line). (lower curve) Plot of the normalized autocorrelation function for a system with the same motor reaction probabilities as in the upper curve, plus a bulk phase reaction with rate constants $k_2 = k_{-2} = 0.0005$ (red circles). These results are compared with the numerical Laplace inversion of Eq.~(\ref{Na_z}) (solid line) discussed in Sec.~\ref{sec:bulk_reaction}. }
\label{Fluct_t}
\end{figure}
The phenomenological rate coefficient is given by $k=\lim_{t \to \infty} k(t)$, and for long times we have the chemical rate law, $dC_{AA}(t)/dt =-(k/V)C_{AA}(t)$, whose domain of validity can be determined from the direct microscopic simulation of $C_{AA}(t)$. This autocorrelation function is plotted in Fig.~\ref{Fluct_t}. Its decay is approximately exponential but, as we shall show below, there are power-law contributions at long times.

\begin{figure}[htbp]
\centering
\resizebox{0.9\columnwidth}{!}{%
\includegraphics{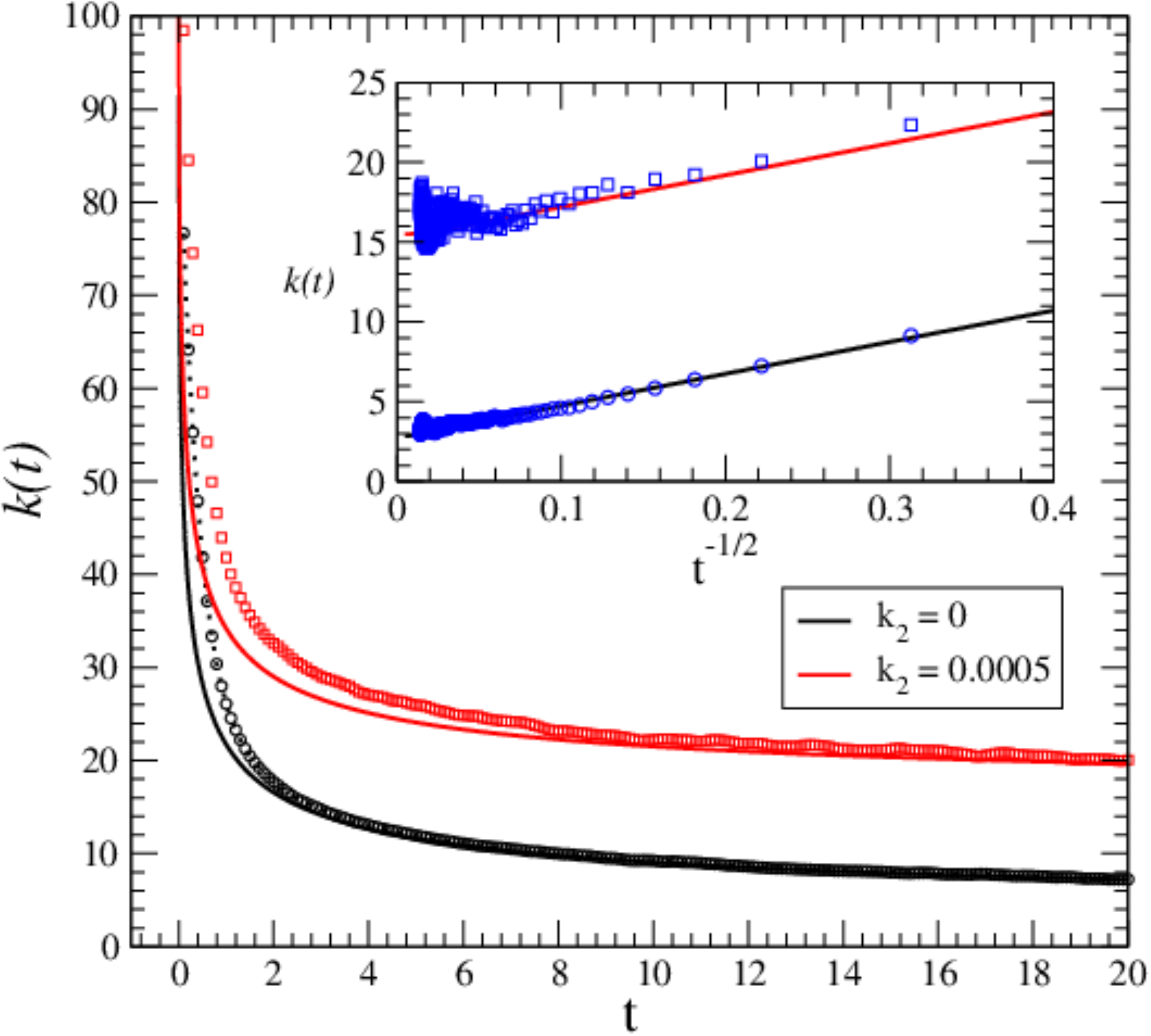}}
\caption{\label{fig:rates} Comparison of simulation and continuum theory results for the integrated rate kernel $k(t)$. The black and red curves correspond to the continuum solution with and without bulk reactions in Eqs.~(\ref{eq:time-rate-coefficient-nb}) and (\ref{eq:time-rate-coefficient}), respectively. The inset shows the long-time asymptotic behavior where the integrated rate kernels approach the long-time value $k$ as $t^{-1/2}$. }
\end{figure}

The reactive dynamics can be probed in more detail by studying the time evolution of the time-dependent rate coefficient $k(t)$. In particular, we now show that the coupling of the reaction at the motor surface to the diffusion of particles leads to a weakly non-exponential, algebraic decay of the number fluctuations that is difficult to detect by visual examination of Fig.~\ref{Fluct_t}. The rate coefficient $k(t)$ can be obtained from the simulation by computing $k(t) \approx -V(dC_{AA}(t)/dt)/C_{AA}(t)$, and the results are plotted in Fig.~\ref{fig:rates}. One sees that $k(t)$ decays very rapidly on a time scale $t_{{\rm mic}} \approx 1$ followed by a weak power-law $t^{-1/2}$ decay (see inset in the figure). Since $t_{{\rm mic}} \ll  t_{{\rm chem}}$ one expects and finds that the phenomenological rate law provides a good approximation to the long-time evolution of $C_{AA}(t)$.

\subsection{Continuum description}\label{subsec:cont}

In the continuum description of the chemical rate processes we again suppose that the system is initially displaced from chemical equilibrium by a small amount but compute the decay to equilibrium by solving the deterministic reaction-diffusion equations. The local concentrations of species $\alpha=A,B$ satisfy the diffusion equation,
\begin{equation}
\frac{\partial c_\alpha ( \bm{r},t) }{\partial t} = D \nabla^2 c_\alpha(\bm{r},t),
\label{diffusion}
\end{equation}
where $D$ is the common diffusion constant of the fluid particles. This equation must be solved subject to the radiation boundary condition~\cite{Collins_Kimball_49} at $r=R$,
\begin{equation}
D \hat{\bm{n}}\cdot \bm{\nabla} c_\alpha(r,\theta,t)\big|_R = -\nu_\alpha\frac{k^{0}}{4\pi R^2}  \psi(R,\theta,t) \Theta(\theta),
\label{eq:rad_bc_1}
\end{equation}
where $k^0=k_+^0+k_-^0$, the stoichiometric coefficients are $\nu_A=-1$ and $\nu_B=1$, $\psi=(k_+^0c_A-k_-^0 c_B)/k^0$ and $\Theta(\theta)$ is the characteristic function that is unity on the catalytic hemisphere ($0 < \theta < \pi /2$) and zero on the noncatalytic hemisphere ($\pi/2 < \theta < \pi$).

Equation~(\ref{diffusion}) can be integrated over the volume of the system outside of the Janus particle with radius $R$ to obtain an evolution equation for $N_A(t)$. Using the boundary condition in Eq.~(\ref{eq:rad_bc_1}), the result of this integration can be written as
\begin{equation}
 \frac{d \, \delta N_A(t)}{dt} =   - k^0  \; \oline{\psi(R,\theta, t)}^{\rm s},
\label{number_A-nb}
\end{equation}
where $\oline{\psi(\bm{r},t)}^{\rm s} = (4\pi R^2)^{-1}\int dS \, \psi(R,\theta, t) \Theta (\theta)$ is the surface average over the catalytic hemisphere at radial distance $R$. The Laplace transform of this equation is
\begin{equation}
z \delta \widehat{N}_A(z) - \delta N_A(0)=- k^0  \; \oline{\widehat{\psi}(R,\theta, z)}^{\rm s}
\end{equation}
with $\delta N_A(0)=\delta N_A(t=0)$. From a knowledge of $\widehat{\psi}(\bm{r}, z)$ given in Appendix~\ref{app:sol_rd_eq}, $\delta \widehat{N}_A(z)$ may be computed and is
\begin{equation}\label{Na_z-nb}
\delta \widehat{N}_A(z) = \frac{V\psi(0)}{z} \left[1 - \frac{a_0(z)}{V} \frac{k^0(1+\nu_0(z) R)}{z}\right] ,
\end{equation}
where $\nu_0^2(z) = z /D$ and $\psi(0)= \delta N_A(0)/V$. An expression for $a_0(z)$ is also given in Appendix~\ref{app:sol_rd_eq}.
The quantity $\delta N_A(t)$ may then be obtained by numerical Laplace inversion.

Rearranging the Laplace transform of the generalized rate law~(\ref{eq:rate-law-mem}) we can write the Laplace transform of the time-dependent rate coefficient as
\begin{align}
\frac{\hat{k}(z)}{V} = \frac{1}{V}\frac{\hat{\phi}_k(z)}{z}
= \frac{\delta N_A(0)}{z \delta \widehat{N}_A(z)} - 1.
\label{eq:rate-kz-nb}
\end{align}
Inserting the solution for $\delta \widehat{N}_A(z)$ into Eq.~(\ref{eq:rate-kz-nb}), we find that
\begin{equation}\label{eq:time-rate-coefficient-nb}
\hat{k}(z) =\frac{ V a_0(z)k^0 (1 + \nu_0 (z) R)}{zV - a_0(z) k^0 (1 + \nu_0 (z) R)}.
\end{equation}
From this equation the short time limit of the rate coefficient is given by $k(t=0^{+}) = k^0/2$ since $\lim_{z \rightarrow \infty} \nu_0(z) R \, a_0(z) = 1/2$.

After numerical Laplace inversion the results of these solutions are plotted in Fig.~\ref{Fluct_t} (the upper curve) where they are compared with the microscopic simulation results for equilibrium systems. Good agreement is obtained. The time dependent rate coefficient $k(t)$ is shown in Fig.~\ref{fig:rates}. The results are close to those from the microscopic simulations but there are observable differences at short times. The inset compares the long-time behavior and shows the $t^{-1/2}$ decay which has its origin in the coupling of the reaction rate to the diffusive motions of the solute species.~\cite{kapral:81}

\section{Janus motor dynamics out of equilibrium}\label{sec:nonequib}
Thus far we have considered a reactive Janus particle in a system at equilibrium where self-propulsion is not possible; however, if the system is driven out of equilibrium by fluxes of reactive species into and out of the system the Janus particle can act as a self-propelled motor that operates by a diffusiophoretic mechanism. Specifically, the system is maintained in a nonequilibrium steady state by contact with reservoirs containing a solution with constant concentrations $\bar{c}_\alpha$ of the chemical species $\alpha$. The reservoirs serve to fix the concentrations at $\bar{c}_\alpha$ at distances far from the Janus particle.

The continuum description of Janus propulsion for this case is well known.~\cite{A89,ALP82,AP91,AB06} From a fluctuating chemohydrodynamics perspective the overdamped motion of the Janus motor is governed by the Langevin equation~\cite{RBELS12,GK17,GK18a},
\begin{equation}
\frac{d{\bf r}_{J}}{dt} = {\bf V}_{\rm d}  + {\bf V}_{\rm fl}(t) \, .
\label{eq-r}
\end{equation}
(The inclusion of an external force ${\bf F}_{\rm ext}$ acting on the motor will be considered in Sec.~\ref{sec:extF} below.) In this equation the fluctuating velocity ${\bf V}_{\rm fl}(t)$ has zero mean, $\langle {\bf V}_{\rm fl}(t)\rangle = 0$, and satisfies the fluctuation-dissipation relation,
\begin{equation}
\langle {\bf V}_{\rm fl}(t) {\bf V}_{\rm fl}(t')\rangle = 2D_t \, \delta(t-t')\, {\boldsymbol{\mathsf 1}},
\end{equation}
where $D_t$ is the translational diffusion coefficient of the motor. The diffusiophoretic velocity ${\bf V}_{\rm d}$ is given by~\cite{GK18a}
\begin{equation} \label{Vd}
{\bf V}_{\rm d} =\frac{1}{1+2b/R} \sum_{\alpha =A}^B  b_{\alpha} \, \overline{\pmb{\nabla}_\perp c_\alpha({\bf r})}^{\rm s} =  V_{\rm d} \, \hat{\bf u},
\end{equation}
where $\pmb{\nabla}_\perp$ stands for the tangential surface gradient and the overline indicates an average over the surface of the Janus particle. The expression for the diffusiophoretic velocity is written for the case of arbitrary slip with a slip coefficient $b$ and diffusiophoretic constants~\cite{AB06,GK18a} $b_\alpha$, where
\begin{equation}
b_\alpha = \frac{k_{\rm B}T}{\eta} \left( K_\alpha^{(1)} + b\, K_\alpha^{(0)}\right) \, ,
\label{b_k}
\end{equation}
with
\begin{equation}
K_\alpha^{(n)} \equiv \int_R^{R+\delta} dr \, (r-R)^n \, [ {\rm e}^{-\beta u_\alpha(r)}-1] \, , \label{K-dfn}
\end{equation}
and $\delta$ is the finite range of the radial intermolecular potentials $u_\alpha(r)$.

The steady-state concentration fields that enter the expression for the diffusiophoretic velocity can be obtained by solving the diffusion equations $\nabla^2 c_\alpha(r,\theta) = 0$ subject to the boundary conditions, $c_\alpha(r=R_{{\rm m}}) = \bar{c}_\alpha$, where $R_{{\rm m}}$ is a distance far from the Janus particle, and the radiation boundary condition on the motor reactive surface at $r=R$ (see Eq.~(\ref{eq:rad_bc_1})). The solution can be written as a series of Legendre polynomials,
\begin{equation}
c_\alpha(r,\theta) = \bar{c}_\alpha + \nu_\alpha (k_+^0 \bar{c}_A - k_-^0 \bar{c}_B) \frac{1}{k_D}\sum_{\ell=0}^{\infty} a_{\ell} f_{\ell} P_{\ell}(\mu),
\label{eq:conti_cB_L}
\end{equation}
where $k_D=4\pi RD$, $\mu = \cos\theta$ and the radial function $f_{\ell}(r) = (R/r)^{\ell+1} - (R/R_{{\rm m}})^{\ell+1}(r/R_{{\rm m}})^{\ell} $. Since the functions $f_{\ell}(R_{{\rm m}}) = 0$, we have $c_\alpha(R_{{\rm m}},\theta) = \bar{c}_\alpha$. The $a_{\ell}$ coefficients can be obtained by solving a set of linear equations,
$M_{\ell m} = G_{\ell m } + (k_+^0 + k_-^0)k_D^{-1}[1-(R/R_{{\rm m}})^{2\ell +1}] K_{\ell m}$ with $G_{\ell m } = [2(\ell+1)/(2\ell+1) + 2\ell/(2\ell+1) (R/R_{{\rm m}})^{2\ell+1}] \delta_{\ell m}$ and $K_{\ell m} = \int_0^1 P_{\ell}(\mu) P_{m}(\mu) d\mu $.
\begin{eqnarray}
&&a_{\ell} = \sum_{\ell = 0}^{\infty} (\mathbf{M})_{\ell m}^{-1} E_m, \\
&&M_{\ell m} = G_{\ell m } + (k_+^0 + k_-^0)k_D^{-1}[1-(R/R_{{\rm m}})^{2\ell +1}] K_{\ell m}, \nonumber\\
&&E_{m} = \int_0^1 P_m(\mu) d\mu,\quad K_{\ell m} = \int_0^1 P_{\ell}(\mu) P_{m}(\mu) d\mu, \nonumber \\
&&G_{\ell m } = [2(\ell+1)/(2\ell+1) + 2\ell/(2\ell+1) (R/R_{{\rm m}})^{2\ell+1}] \delta_{\ell m}.\nonumber
\end{eqnarray}
Explicit expressions for $E_m$ and $K_{\ell m}$ are given in Eqs.~(\ref{eq:Mmatrix}) and (\ref{eq:Evector}) of Appendix~\ref{app:sol_rd_eq}.

Substituting these expressions for the concentration fields into Eq.~(\ref{Vd}) we obtain an expression for the diffusiophoretic velocity of the Janus particle,
\begin{eqnarray}\label{eq:Vd-cont}
V_{\rm d} &=& \frac{2k_{\rm B}T}{3\eta}\frac{(\Lambda^{(1)} +b \Lambda^{(0)})}{2b+R} \\
&&\times \frac{(k_+^0 \bar{c}_A - k_-^0 \bar{c}_B)}{k_D}[ 1 - (R/R_{{\rm m}})^3]a_1,\nonumber
\end{eqnarray}
where we have defined $\Lambda^{(n)}=K_B^{(n)}-K_A^{(n)}$.

\subsection{Simulation of nonequilibrium Janus dynamics}

We now compare microscopic simulations with the Langevin model derived from nonequilibrium fluctuating thermodynamics. The nonequilibrium steady state conditions discussed above can be implemented in the microscopic simulations as follows. Consider a spherical region with radius $r=R_{{\rm m}}=24$ centered on the Janus motor. Whenever a fluid particle enters the region $r \le R_{{\rm m}}$ its species label is changed to $\alpha=A,B,S$ with probability $\bar{p}_{\alpha}$. The resulting concentration of species $\alpha$ at $r=R_{{\rm m}}$ is $\bar{c}_{\alpha} = c_0 \bar{p}_{\alpha}$, where $\sum_{\alpha} \bar{p}_{\alpha} = 1$ and $c_0 = \sum_{\alpha} \bar{c}_{\alpha}$ is the fixed total number of particles per unit volume. This simulates a system where the concentrations outside of the $r=R_{{\rm m}}$ boundary are prescribed to be $\bar{c}_{\alpha}$ for $\alpha$-type particles.

In the simulations we consider a system with $c_0 =20$ and repulsive interaction strengths $\epsilon_A = 1$, $\epsilon_B = 0.1$ and $\epsilon_S = 0.5$. The reversible Janus catalytic reactions use $p_{\pm} = 1$ so that $k_{\pm}^0 = p_{\pm}\nu_c^0 = 188.4$. To implement the nonequilibrium boundary conditions we take $\bar{p}_A = 0.5$ and $\bar{p}_B = 0.45$. With these parameters, the system deviates slightly from equilibrium so that it remains in the linear regime. However, since $k_+^0/k_-^0 =p_+/p_-\neq \bar{p}_B/\bar{p}_A$ detailed balance is broken and motor self-propulsion can occur. 

In order to evaluate Eq.~(\ref{eq:Vd-cont}) for the diffusiophoretic velocity we require various input parameters. The $\Lambda^{(n)}$ factors in Eq.~(\ref{eq:Vd-cont}) involve the interaction potentials $u_\alpha(r)$ that can be identified as angular averages of the potentials of mean force defined in Eq.~(\ref{eq:reduced-den}). Consequently, these factors can be computed from a knowledge of the radial distribution functions $g_{\alpha}(r)=e^{-\beta u_{\alpha}(r)}$,
\begin{equation}
\Lambda^{(n)} = \int_0^{\infty} dr\:r^n [g_{B}(r)-g_{A}(r) ] ,
\end{equation}
where the integrals may be extended over all $r$ values since the integrand vanishes in the interior of the Janus particle and outside the range of the mean potential. For a system with $\epsilon_A=1.0$ and $\epsilon_B=0.1$ we have $\Lambda^{(0)}=0.1006$ and $\Lambda^{(1)}=0.4798$. The values of the solute diffusion coefficient $D$ and fluid viscosity $\eta$ are given in Appendix~\ref{app:sim}, and the solution of the reaction-diffusion equation yields $a_1 = 5.25\times 10^{-3}$.

The remaining parameter to determine is the slip length $b$. To estimate this quantity we assume that the translational and rotational diffusion coefficients have their hydrodynamic values,
\begin{equation}
D_t = \frac{k_{\rm B}T}{6\pi\eta R} \frac{1+3b/R}{1+2b/R}, \quad D_r = \frac{k_{\rm B}T}{8\pi\eta R^3}(1+3b/R),
\end{equation}
and equate them to the simulation values of these transport coefficients. From the mean square displacement we obtain $D_t = 9\times 10^{-4}$, while decay of the orientational correlation function, $C_{\rm u}(t) =\langle \hat{{\bf u}}(t)\cdot \hat{{\bf u}} \rangle = \exp(-2D_r t)$ and yields $D_r = 1.37\times 10^{-4}$.
Given that $R=5$, $\eta=16.58$, and $k_{\rm B}T=1$ in the chosen units, we find $b \simeq 11$ from $D_t$ and $b \simeq 10$ from $D_r$. Using $b=10.5$ and the other parameter values in Eq.~(\ref{eq:Vd-cont}) we find the theoretical estimate $V_{\rm d} = 6.2\times 10^{-4}$, which is comparable to the simulation result $V_{\rm d}= 6.0\times 10^{-4} \pm 2\times 10^{-5}$. The relatively large value of the slip length $b$ finds its origin in the slipperiness of the fluid-particle interface due to the repulsive potentials used in the simulations.

The microscopic simulation of the autocorrelation function of the fluctuating velocity, $\langle {\bf V}_{\rm fl}(t) {\bf V}_{\rm fl}(0)\rangle$, was also computed. It exhibits a rapid decay on a time scale of $t_{{\rm v}} \approx 0.7$, with a power-law tail at longer times. The time integral of this correlation function gives $D_t = 10^{-3}$, which is consistent with the value obtained from the mean square displacement.

We may also consider microscopic aspects of the reaction rate. From fluctuating thermodynamics the reaction rate, $dn/dt$, gives the instantaneous time rate of change of the net number of product molecules that are produced in the motor catalytic reaction up to time $t$. It is a fluctuating random variable that satisfies the stochastic equation,~\cite{GK17,GK18a}
\begin{equation}
\frac{dn}{dt} = W_{\rm rxn} + W_{\rm fl}(t) \, ,
\label{eq-n}
\end{equation}
where $W_{\rm rxn}$ is the mean reaction rate and $W_{\rm fl}(t)$ is the fluctuating rate that satisfies the fluctuation-dissipation relation, $\langle W_{\rm fl}(t)\, W_{\rm fl}(t')\rangle = 2D_{\rm rxn} \, \delta(t-t') $, with $D_{\rm rxn}$ the reaction diffusivity.

The mean reaction rate gives the average value of the rate at which product molecules are produced, $W_{\rm rxn} = W_+ -W_-$, where $W_\pm$ are the rates of the forward and reverse reactions. It is zero in equilibrium but takes non-zero values under nonequilibrium conditions. Since $W_+ = \int_S dS\:k_+^0\:c_A(R,\theta)\:\Theta(\theta)$ and $W_- = \int_S dS\:k_-^0\:c_B(R,\theta)\:\Theta(\theta)$, where the surface integrals are restricted to the motor catalytic surface by the characteristic function $\Theta(\theta)$, using the expressions for $c_\alpha(R,\theta)$ in Eq.~(\ref{eq:conti_cB_L}), we find $W_{\rm rxn} = \Gamma (k_+^0 \bar{c}_A - k_-^0 \bar{c}_B)$, where $\Gamma = (1 - \gamma_J k^0 /k_D )/2$ and $\gamma_J = \sum_{\ell = 0}^{\infty} a_{\ell} [1 - (R/R_{\rm m})^{2\ell +1}] E_\ell$. The reaction diffusivity is given by $D_{\rm rxn}=(W_+ + W_-)/2 =\Gamma (k_+^0 \bar{c}_A + k_-^0 \bar{c}_B)/2$. For our system parameters, $\Gamma = 0.0085$ with $\gamma_J = 0.0097$ and $k_+^0 \bar{c}_A - k_-^0 \bar{c}_B = 188.4$ so that $W_{\rm rxn} = 1.6$ and $D_{\rm rxn} = 15.3$.

The Fokker-Planck equation for the probability $p(n;t)$ that $n$ product molecules have been produced up to time $t$  corresponding to the Langevin equation~(\ref{eq-n})  is
\begin{equation}
\frac{\partial p}{\partial t}= -W_{\rm rxn} \partial_n p +D_{\rm rxn} \partial^2_n p,
\end{equation}
whose solution is
\begin{equation}
p(n;t) = \frac{1}{\sqrt{4\pi D_{\rm rxn}t  }} \exp{\bigg[ \frac{-(n - W_{\rm rxn} t )^2}{ 4 D_{\rm rxn} t}\bigg]}.
\label{eq:Pn}
\end{equation}

\begin{figure}[htbp]
\centering
\resizebox{1.0\columnwidth}{!}{%
\includegraphics{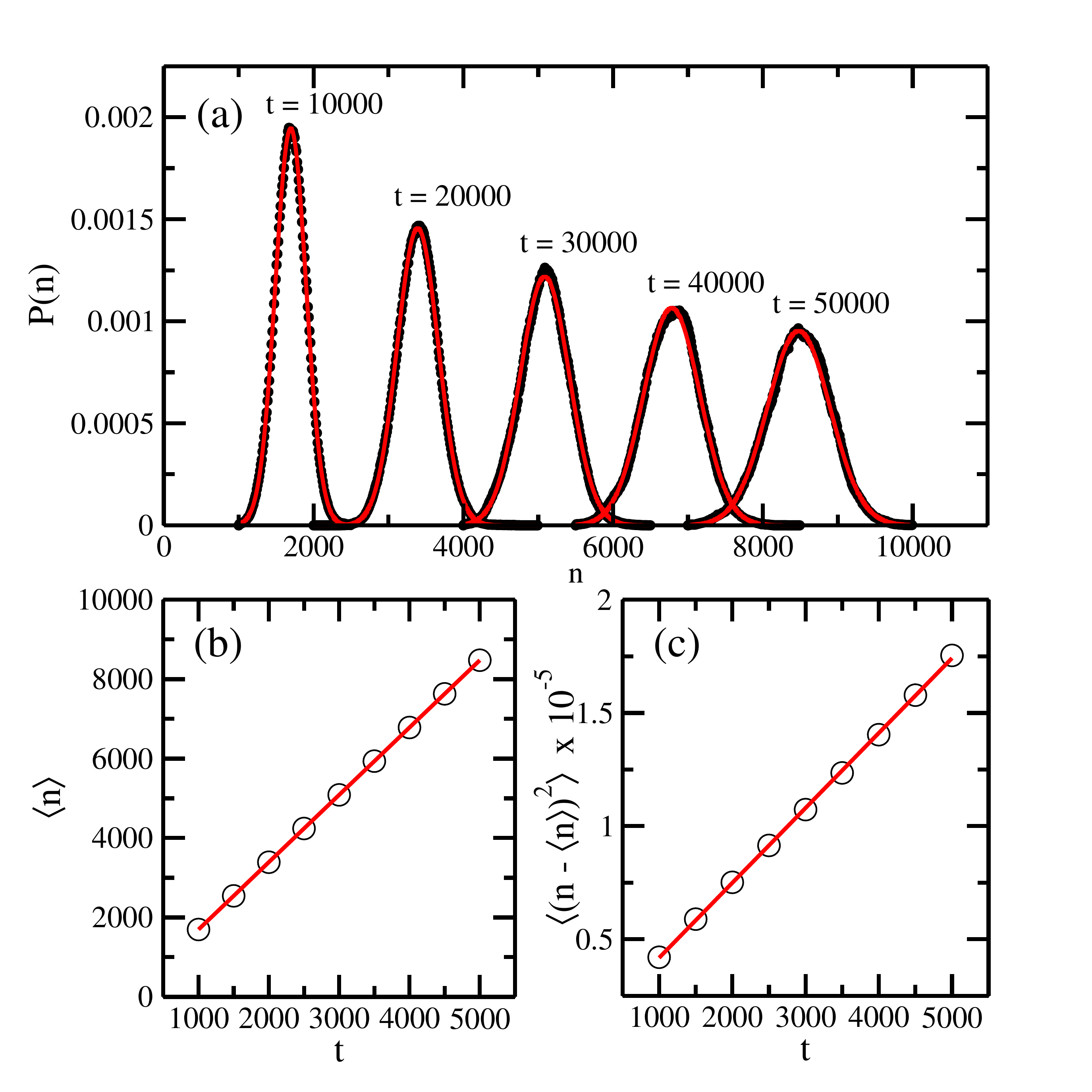}}
\caption{\label{fig:p_n}  (a) The probability distribution functions, $p(n;t)$, at different times, where the black dots are simulation results and red curves are the Gaussian  distribution (Eq.~(\ref{eq:Pn})) with simulation values of mean and variance shown in panels (b) and (c), respectively.  }
\end{figure}

The long-time steady-state values of $W_{\rm rxn}$ and $D_{\rm rxn}$ can be determined from the  distribution of product particles estimated from simulations. Figure~\ref{fig:p_n}(a) shows the probability distributions obtained by constructing histograms of $n$. The results are consistent with Gaussian distributions (red curves) with the mean and variance of $n$ shown in panels (b) and (c), respectively. From these results we find $W_{\rm rxn} = 1.7$ and $D_{\rm rxn} = 16.5$, in good agreement with the continuum theory results $1.6$ and $15.3$, respectively.

\subsection{Janus motor subject to an external force}\label{sec:extF}

In the presence of an external force ${\bf F}_{\rm ext}$ one must consider the coupled Langevin equations~\cite{GK17,GK18a},
\begin{eqnarray}
\frac{d{\bf r}_{J}}{dt} &=& {\bf V}_{\rm d}  + \beta D_t \, {\bf F}_{\rm ext} + {\bf V}_{\rm fl}(t) \, ,\label{eq-LwF}\\
\frac{dn}{dt} &=& W_{\rm rxn} + \beta\chi D_{\rm rxn} \hat{\bf u}\cdot{\bf F}_{\rm ext} +W_{\rm fl}(t) \, ,\label{eq-nF}
\end{eqnarray}
in order for the fluctuating thermodynamics description to be consistent with microscopic reversibility. In particular, this consistency requires that a contribution, $W_{\rm d}= \beta\chi D_{\rm rxn} \hat{\bf u}\cdot{\bf F}_{\rm ext}$, that is reciprocal to the diffusiophoretic coupling appear in Eq.~(\ref{eq-nF}). Here $\chi = V_{\rm d}/ W_{\rm rxn} $. As a result of this contribution the reaction rate depends on the external force, and allows for the possibility that the application of an external force can result in the net product of fuel from product.

To investigate the consequences of this reciprocal contribution on the reaction rate, in the microscopic simulations we subject the Janus particle with a magnetic moment $\mu$ to an external force and torque that are derived from the external potential function $U_{\rm ext}(\mathbf{r}_{J},\hat{\mathbf{u}})  = -\mathbf{F}_{\rm ext}\cdot \mathbf{r}_{J}  - \mu \mathbf{B}\cdot \hat{\mathbf{u}}$, where $\mathbf{F}_{\rm ext} = F_{\rm ext} \hat{\mathbf{z}}$ and $\mathbf{B} = B\:\hat{\mathbf{z}}$ is the external magnetic field chosen to be in the same direction as external force. The magnetic field produces an external torque $\mathbf{T}_{\rm ext} = \mu\:\hat{\mathbf{u}} \times \mathbf{B}$ that tends to align the Janus motor with $\mathbf{B}$ and thus with the external force. (See Appendix~\ref{app:sim} for details.)

\begin{figure}[htbp]
\centering
\resizebox{1.0\columnwidth}{!}{%
\includegraphics{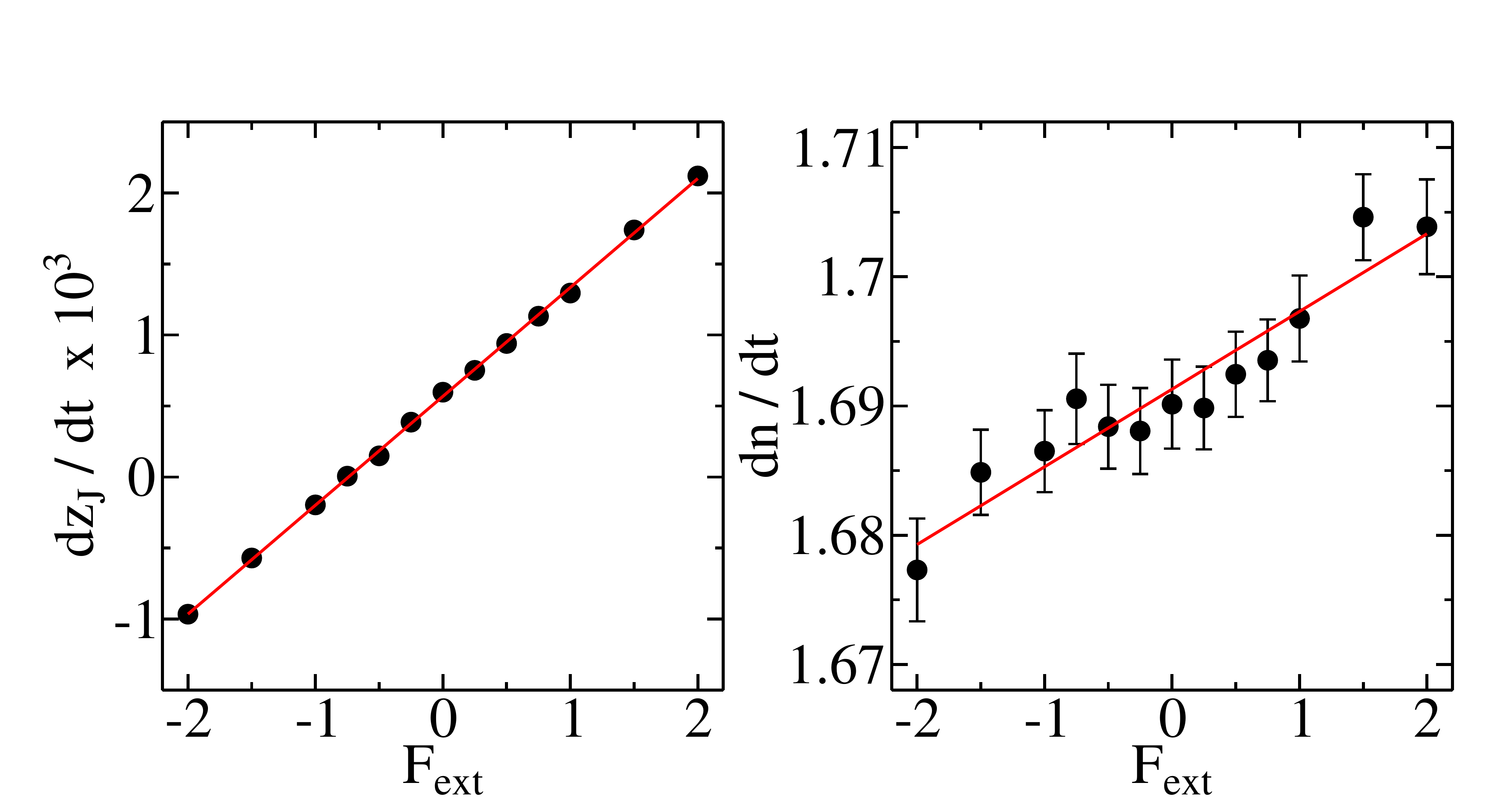}}
\caption{\label{fig:n_t} Plots of the $F_{\rm{ext}}$ dependence of the average motor velocity in the $\hat{\bm{z}}$ direction, $d\langle z_J \rangle/dt$ (left panel), and of the reaction rate, $d\langle n \rangle/ dt$ (right panel). The fits to the data given in the text are indicated by red lines. The results were obtained from averages over 200 realizations of the dynamics.}
\end{figure}

In the simulation we can compute the average Janus velocity and reaction rate and compare the results with the averages of Eqs.~(\ref{eq-LwF}) and (\ref{eq-nF}),
\begin{eqnarray}
 \frac{d  \langle z_J\rangle }{dt} &=& \chi W_{\rm rxn}\langle \hat{u}_z\rangle + \beta D {F}_{\rm ext},\\
\frac{d \langle n \rangle}{dt} &=&  W_{\rm rxn} + \beta \chi D_{\rm rxn} \langle \hat{u}_z\rangle {F}_{\rm ext},
\end{eqnarray}
where $z_J = {\bf r}_J \cdot \hat{\bf z}$.
Simulations were carried out with a magnetic field strength $B=500$ with a magnetic moment $\mu=1$. In this case the simulation yields $\langle \hat{u}_z\rangle =0.998$ which agrees with the the theoretical estimate $\langle \hat{u}_z\rangle={\rm coth}(\beta\mu B)-1/(\beta\mu B) = 0.998$. The simulation results for $d  \langle z_{J} \rangle /dt$ and $d \langle n \rangle/dt$ versus ${F}_{\rm ext}$ are plotted in Fig.~\ref{fig:n_t}.

One can see that both the projected motor velocity and the average reaction rate increase linearly with the external force. Fits to these data yield $d\langle z_J \rangle/dt = 5.7\times10^{-4} + 7.7\times10^{-4} {F}_{\rm ext}$ and $d\langle n \rangle/ dt = 1.69 + 6.0 \times 10^{-3} {F}_{\rm ext}$. The linear fit to the simulation data for $d\langle n \rangle/ dt$ is found to have slope $6.0 \times 10^{-3}$, which agrees with the theoretical prediction of $\beta \chi D_{\rm rxn} \langle \hat{u}_z\rangle \simeq 6\times 10^{-3}$, confirming the existence of the effect of the external force on the production rate of product particles on the motor surface due to the diffusiophoretic coupling.

\section{Inclusion of a fluid phase reaction}\label{sec:bulk_reaction}

While the catalytic cap on the Janus motor catalyzes the reaction $A +C \rightleftharpoons B+C$, it is possible that this reaction can also take place in the bulk fluid phase in the absence of catalyst. Here we suppose that this is possible and include the fluid phase reaction $A \underset{k_{-2}}{\stackrel{k_2}{\rightleftharpoons}} B$. Since a catalyst does not alter the equilibrium in the system and only changes the forward and reverse rate constants, in order to satisfy detailed balance we must have $k^0_+/k^0_-=k_2/k_{-2}= K_{{\rm eq}}$.

When a bulk phase reaction is also present the generalized rate law has a form analogous to that in Eq.~(\ref{eq:rate-law-mem-C}), and can be written as
\begin{equation}\label{eq:rate-law-mem-bulk}
\frac{d}{dt}C_{AA}(t) =- (k_2 + k_{-2}) C_{AA}(t) - \int_{0}^{t} d\tau \; \frac{\phi_k(\tau )}{V} C_{AA}(t-\tau),
\end{equation}
with a modified time dependent rate coefficient that includes the bulk reaction and is defined by
\begin{equation}
\frac{k(t)}{V} = k_{2} + k_{-2} + \frac{1}{V} \int_0^t d\tau \; \phi_k(\tau).
\end{equation}
In the simulation the fluid phase reaction is taken into account by using reactive multiparticle collision dynamics.~\cite{Rohlf_etal_08} The dynamics preserves the conservation laws and detailed balance. Additional details are given in Appendix~\ref{app:sim}. The results of simulations of $C_{AA}(t)$ and $k(t)$ for an equilibrium system with both motor catalyzed reactions and uncatalyzed fluid phase reactions are shown in Figs.~\ref{Fluct_t} and \ref{fig:rates}. The structures of these functions are similar to those of systems where no fluid phase reaction is present, although, as expected, the decay is more rapid because of the increased bulk phase reactivity of the system.

\subsection{Continuum description with fluid phase reaction}

We may again compare the microscopic simulation results with those predicted from a continuum model. The continuum description in Sec.~\ref{subsec:cont} is easily extended to include a fluid phase reaction. As earlier, we suppose that the system is initially displaced from chemical equilibrium by a small amount and compute the decay to equilibrium. The reaction-diffusion equation for $c_\alpha(\mathbf{r},t)$ now takes the form.
\begin{equation}
\frac{\partial c_\alpha ( \bm{r},t) }{\partial t} = D \nabla^2 c_\alpha(\bm{r},t)
+ \nu_\alpha(k_2 c_A(\bm{r},t) - k_{-2} c_B(\bm{r},t)),
\label{reaction-diffusion}
\end{equation} and the reaction at the surface of the motor is accounted for through the radiation boundary condition in Eq.~(\ref{eq:rad_bc_1}).

Following the earlier derivation, Eq.~(\ref{reaction-diffusion}) can be integrated over the volume of the system outside of the Janus particle with radius $R$ to obtain
\begin{equation}
 \frac{d \, \delta N_A(t)}{dt} =- (k_2 + k_{-2}) \delta N_A(t)- k^0  \; \oline{ \psi(R,\theta, t)}^{\rm s}.
\label{number_A-bulk}
\end{equation}

From the solution for the Laplace transform of $\psi(R,\theta, t)$ outlined in Appendix~\ref{app:sol_rd_eq} we may obtain $\delta \widehat{N}_A(z)$ and $\hat{k}(z)$, which are given by
\begin{eqnarray}
\delta \widehat{N}_A(z) &=& \frac{\delta N_A(0)}{D \nu^2(z)} \left[
1 - \frac{a_0(z)}{V} \frac{k^0(1+\nu (z) R)}{D\nu^2(z)}
\right] ,
\label{Na_z}
\end{eqnarray}
where $\nu^2(z) = (z + k_2 +k_{-2})/D$, and
\begin{align}
\hat{k}(z) = &\frac{ V(k_2+k_{-2})}{z} \label{eq:time-rate-coefficient} \\
&+ \frac{D\nu^2(z)}{z} \frac{  V a_0(z) k^0 (1+\nu(z)R)}{V D\nu^2(z) - a_0(z)k^0(1+\nu(z) R)}.
\nonumber
\end{align}
The short-time limit of the rate coefficient is $k(0^{+}) = V(k_2 + k_{-2}) + k^0/2$ since $\lim_{z \rightarrow \infty} \nu (z) R \, a_0(z) = 1/2$.
The results for $\delta N_A(t)$ and $k(t)$ obtained by numerical Laplace inversion are plotted in Figs.~\ref{Fluct_t} and \ref{fig:rates} and agree well with the microscopic simulations at long times but as expected, exhibit notable differences at short times.

\subsection{Nonequilibrium fluid phase reaction}

It is possible that the reactive $A$ and $B$ species may participate in other chemical reactions that are themselves taking place under nonequilibrium conditions. For example suppose that the chemical reaction
$E+A \underset{k_{-3}}{\stackrel{k_3}{\rightleftharpoons}} F+B$ takes place in the fluid phase and that the $E$ and $F$ chemical species are pool species whose concentrations are fixed and may be incorporated into the $k_{\pm 3}$ rate constants so that $k_n= k_3 c_E$ and $k_{-n}= k_3 c_F$. By varying the concentrations of the pool species one can break detailed balance since $k_+^0/k_-^0 \neq k_{n}/k_{-n}$, and maintain the system in a nonequilibrium state.

In a nonequilibrium steady state produced in this way, concentration gradient fields of chemical species will be generated and motor self-propulsion will take place. For example, we have simulated systems with energy parameters $\epsilon_A = 1$, $\epsilon_B = 0.5$ and $\epsilon_S = 0.5$ and motor catalytic reaction probabilities $p_{\pm} = 0.5$ for two choices of $k_{\pm n}$ that break detailed balance.  When (a) $k_n = 10^{-3}$ and $k_{-n} = 10^{-2}$ we find that the diffusiophoretic velocity is $V_{\rm d}=0.0017$, while for the other choice where (b) $k_n= 10^{-2}$, $k_{-n} = 10^{-3}$ we find that the motor moves with the negative velocity $V_{\rm d}=-0.0016$, showing that the motor motion can be controlled by altering how the system is driven out of equilibrium.

\section{Conclusion} \label{sec:conc}

The coarse-grain microscopic model incorporating reversible reaction kinetics on the catalytic face of a Janus particle was shown to provide a description of the dynamics that conserves mass, momentum and energy with reactive events that satisfy microscopic reversibility. Consequently, fundamental features of the dynamics of these particles could be investigated. Our results for a reactive Janus particle in an equilibrium system showed that the model is able to capture all of the properties expected in such a system, namely, a binomial distribution of chemical species and a generalized chemical rate law with a time-dependent rate coefficient that has a long-time power law decay due to coupling of reaction to solute diffusion modes.

We also showed that when the system is driven out of equilibrium by coupling it to reservoirs with constant concentrations of chemical species, or by out-of equilibrium fluid phase reactions, detailed balance is broken and the Janus particle can become a motor and move autonomously by self-diffusiophoresis. The results of the microscopic model were compared with deterministic and stochastic theories based on continuum reaction-diffusion and hydrodynamic equations of motion. In particular, when the Janus motor is subject to an external force, we were able to verify the existence of an effect that is reciprocal to diffusiophoresis that causes the reaction rate to depend on the external force.

Our study has served to document that the theoretical underpinnings of the microscopic model accurately describe the dynamics of Janus particles under both equilibrium and nonequilibrium conditions. It also showed how the model can be used to complement and extend the predictions of phenomenological theories. The microscopic model can be extended in various ways; for example, by changing the geometry of the motor, implementing other motor and fluid phase reaction mechanisms, and environmental and boundary conditions. As such, the microscopic framework presented here, which satisfies the basic principles of dynamics, can be applied to other active systems.

\section*{Acknowledgments}

Financial support from the International Solvay Institutes for Physics and Chemistry, the Universit\'e libre de Bruxelles (ULB), the Fonds de la Recherche Scientifique~-~FNRS under the Grant PDR~T.0094.16 for the project ``SYMSTATPHYS", and the Natural Sciences and Engineering Research Council of Canada is acknowledged.


\appendix

\section{Simulation method and parameters}\label{app:sim}
The Janus motor is made from $N_b = 2681$ motor beads, each with mass $m$ and radius $\sigma$, randomly distributed within a sphere of radius $R_J = 4\:\sigma$. The effective radius of the Janus motor is $R = R_J + \sigma=5\sigma$. To ensure spherical symmetry, the equilibrium coordinates of the $N_b$ beads are chosen such that the diagonal elements of the moment of inertia tensor are approximately $I = \frac{2}{5} m_J R_J^2$ with small off-diagonal elements, where $m_J = m\:N_b$ is the total mass. In a selected equilibrium configuration, two beads are linked by a harmonic spring with spring constant $k_s = 50\: k_B T/\sigma^2$ if their separation is less than $2\:\sigma$. The Janus motor is placed in a cubic periodic box of linear size $L$ consisting of $N= N_A + N_B + N_S$ solvent particles, and the average solvent density is $n_0 = N / L^3 \approx 20$. For the simulations that deal with the binomial distribution $L = 20$ and $N = 153417$ with $N_A+N_B = 75368$ and $N_S = 78049$, whereas other simulations described in this paper have a box of size $L=50$ and $N=2488439$. In the simulations with bulk reactions, the total number $A$ and $B$ particles is chosen to be $N_A+N_B = 1244219$ and $N_S = 1244220$ inert solvent $S$ particles with $L=50$. The interaction strengths of the repulsive interactions between motor beads and fluid particles as well as the reaction probabilities for specific simulations are given in the text. The results in the paper are reported in dimensionless units where mass is in units of $m$, length in units of $\sigma$, energies in units of $k_{\rm B}T$ and time in units of $t_0 = \sqrt{m \sigma^2/k_{\rm B}T}$.

Interactions among solvent particles are described by multiparticle collision (MPC) dynamics comprising streaming and collision steps at discrete time intervals $\tau = 0.1\:t_0$. During each collision step, fluid particles are sorted into a grid of cubic cells with linear size $\sigma$. The postcollision velocities of particles $i$ in a cell $\xi$ are given by $\mathbf{v}_i' = \mathbf{V}_{\xi} + \hat{\mathcal{R}} (\mathbf{v}_i - \mathbf{V}_{\xi})$, where $\mathbf{V}_{\xi}$ is the center of mass velocity of particles cell $\xi$ and $\hat{\mathcal{R}}$ is a rotation operator about a random axis by an angle of $120^{\circ}$. In the streaming step, the system evolves by Newton's equations of motion with forces determined from the potential function, $U({\bf r}, \bm{\alpha})$, using a time step of $\delta t = 0.005\:t_0$. The mean free path for MPC is $0.1$ in the simulation.

The common diffusion constant of the solvent particles determined from measurements of the mean-squared displacement is found to be $D = 0.06$ in dimensionless units. Using the MPC expression for the viscosity~\cite{Malevanets_Kapral_99,Malevanets_Kapral_00,Kapral_08,gompper:09} one gets $\eta = 16.58$, the kinematic viscosity of the fluid is $\nu = \eta/c_0 = 0.829$ and the Schmidt number is ${\rm Sc} = \nu/D \simeq 14$.

In the simulations with an external force and torque in Sec.~\ref{sec:extF} a thermostat and an effective no-slip boundary are needed in order to keep the system temperature constant and avoid a systematic drift of the entire system in a periodic simulation box. Specifically, in the MPC collision steps the velocities of the particles outside $r=R_{\rm m}$ are drawn from a Maxwell-Boltzmann distribution with zero mean velocity and variance $\sqrt{k_{\rm B}T/m}$. In this way, the system inside $r=R_{\rm m}$ is effectively in contact with a heat bath with temperature $T$ with a vanishing average velocity at the boundary at $r=R_{\rm m}$.

In the simulations with chemical reactions in the fluid phase, reactive multiparticle collision dynamics~\cite{Rohlf_etal_08} was employed and a bulk reaction, $A\underset{k_{-2}}{\stackrel{k_2}{\rightleftharpoons}} B$ was introduced, with $k_2$ and $k_{-2}$ the forward and reverse rate constants, respectively. The fluid phase reactions are carried out at the multiparticle collision steps, where forward and reverse reactions take place independently in each cell with probabilities $p_2^{\xi} = q_2^{\xi}(1 - e^{-q_2^{\xi}})/q_0$ and $p_{-2}^{\xi} = q_{-2}^{\xi}(1 - e^{-q_{-2}^{\xi}})/q_0$, respectively. Here $q_0 = q_2^{\xi} + q_{-2}^{\xi}$ with $q_2^{\xi} = k_2 N_B^{\xi}$ and $q_{-2}^{\xi} = k_{-2} N_A^{\xi}$, where $N_A^{\xi}$ and $N_B^{\xi}$ are the total number of $A$ and $B$ particles in cell $\xi$.

\section{Solution of reaction-diffusion equation}\label{app:sol_rd_eq}
In this Appendix we present the solution of the evolution equations for the concentration fields of the $A$ and $B$ species in a  reaction-diffusion system with a fluid phase reaction $A\underset{k_{-2}}{\stackrel{k_2}{\rightleftharpoons}} B$, as well as a catalytic reaction on the motor surface. We restrict ourselves to systems where detailed balance is satisfied so that $k_+^0/k_-^0=K_{{\rm eq}}$, and if uncatalyzed fluid phase reactions are present, $k_2/k_{-2}=k_+^0/k_-^0=K_{{\rm eq}}$.
The results in Sec.~\ref{subsec:cont} for a system with no fluid phase reaction may be obtained from the solutions in this Appendix by setting $k_2=k_{-2}=0$.

Including the fluid phase reaction, the reaction diffusion equation for $c_\alpha$ ($\alpha=A,B$) is given by Eq.~(\ref{reaction-diffusion}) and this equation must be solved subject to the radiation boundary condition~(\ref{eq:rad_bc_1}). Making the change of variables,
\begin{equation}
c=c_A+c_B, \quad \psi=(k_+^0c_A-k_-^0c_B)/k^0,
\end{equation}
the coupled reaction-diffusion equations and their boundary conditions take the uncoupled forms,
\begin{equation} \label{eq:c}
\frac{\partial c(\bm{r},t)}{\partial t} = D \nabla^2 c(\bm{r},t),\quad
D\hat{\bm{n}}\cdot \bm{\nabla} c(\bm{r},t)\big|_R = 0,
\end{equation}
and
\begin{eqnarray}\label{eq:psi}
&&\frac{\partial \psi(\bm{r},t)}{\partial t} = D \nabla^2 \psi(\bm{r},t)-(k_2+k_{-2})\psi(\bm{r},t),\\
&&\quad D\hat{\bm{n}}\cdot \bm{\nabla} \psi(\bm{r},t)\big|_R = \frac{k^0}{4 \pi R^2}\psi(R,\theta,t)\Theta(\theta).\nonumber
\end{eqnarray}
Note that if the change of variables $c_A=\psi+k_-^0c/k^0$ and $c_B=-\psi +k_+^0c/k^0$ is substituted into the right side of the reaction-diffusion equation~(\ref{reaction-diffusion}), we obtain
\begin{eqnarray}
&&k_2 c_A(\bm{r},t) - k_{-2}^0 c_B(\bm{r},t) =(k_2+k_{-2})\psi \\
&&\qquad \qquad -c(k_+^0/k_-^0-k_2/k_{-2})k_-^0 k_{-2}/k^0;\nonumber
\end{eqnarray}
hence, it is only when the system satisfies detailed balance that the equations and their boundary conditions decouple in the new variables.

We are interested in the solutions of these equations for a system that is initially slightly displaced from chemical equilibrium so that $\delta N_A(0) = N_A(0) - N_A^{\rm eq}=-\delta N_B(0)$. In terms of the new variables we can write,
\begin{equation}
\delta N_A(t)=\int_R d\bm{r} \; \delta c_A(\bm{r},t)= \int_R d\bm{r} \; \psi(\bm{r},t)=-\delta N_B(t),
\end{equation}
where the integrals are over the volume outside of the Janus particle; thus, the information needed to compute these quantities can be obtained from a knowledge of $\psi(\bm{r},t)$.

The Laplace transform of Eq.~(\ref{eq:psi}) is
\begin{align}
\big( \nabla^2 - \nu^2 (z) \big) \, \hat{\psi}(\bm{r},z) = - \frac{1}{D} \psi(\bm{r},0),
\end{align}
where $\nu^2(z) = (z+k_2 + k_{-2})/D$ and $\psi(\bm{r},0)=\delta N_A/V=-\delta N_B/V$, which is independent of $\bm{r}$ for this choice of initial condition. Henceforth we shall not indicate the dependence of the parameter $\nu$ on the Laplace variable $z$. This equation can be solved using the Green function method. The Green function $\hat{g}(\bm{r},\bm{r}^\prime,z)$ for the axisymmetric system satisfies
\begin{equation}
\left( \nabla^2 - \nu^2 \right) \hat{g}(\bm{r},\bm{r}^\prime,z) = - \frac{\delta (r - r^\prime)}{2 \pi r^{\prime 2}}
\sum_{\ell = 0}^\infty \frac{2\ell +1}{2} P_\ell (\mu) P_\ell (\mu^\prime),
\label{eq:Green}
\end{equation}
subject to the radiation boundary conditions specified in Eq.~(\ref{eq:psi}) at the surface of the motor and assuming $\hat{g}(\bm{r},\bm{r}^\prime ,z)$ vanishes far from the Janus motor. In Eq.~(\ref{eq:Green}), $P_\ell(x)$ are the Legendre polynomials, $r$ and $r^\prime$ are radial distances from the center of the Janus particle in a spherical polar coordinate system with polar angles $\theta$ and $\theta^\prime$ measured from a polar axis aligned with the unit vector $\hat{\bm{u}}$  (see Fig.~\ref{fig:Janus}), and $\mu=\cos\theta$ and $\mu^\prime = \cos\theta^\prime$.

A general Green function that vanishes as $r \rightarrow R_m$ and $r' \rightarrow R_m$ can be written in terms of modified spherical Bessel functions $k_{\ell}$ and $i_{\ell}$ that satisfy the radial equation
\begin{align}
\bigg( \frac{d^2}{d r^2} + \frac{2}{r}\frac{d}{d r}  - \left( \frac{\ell (\ell +1)}{r^2} + \nu^2 \right) \bigg) y_{\ell}(\nu r) = 0 .
\end{align}
Exploiting the axial symmetry of the Janus motor system, the Green function can be written in terms of the two independent radial solutions as
\begin{align*}
\hat{g}(&\bm{r},\bm{r}^\prime ,z) = \frac{\nu}{2\pi} \sum_{\ell }^\infty \frac{2\ell + 1}{2} P_\ell (\mu) \bigg(
h_{\ell}(\nu r) i_{\ell}(\nu r') H (r-r') \\
& \qquad + i_{\ell}(\nu r) h_{\ell}(\nu r') H(r' -r) \bigg) P_{\ell}(\mu')\\
&+\frac{\nu}{2\pi} \sum_{\ell, m} \frac{2\ell +1}{2} P_{\ell}(\mu) h_{\ell}(\nu r)
{\Gamma}_{\ell m} h_m(\nu r^\prime) P_m(\mu^\prime) \frac{2m+1}{2}
\end{align*}
where $H(r)$ is the Heaviside function, $h_\ell (r) = k_{\ell}(r) - \alpha_l i_{\ell}(r)$ with $\alpha_l = k_{\ell}(\nu R_m)/i_{\ell}(\nu R_m)$ and $\boldsymbol{\Gamma}$ is a symmetric matrix determined by the radiation boundary conditions.  Note that in the limit that $R_m \rightarrow \infty$, $\alpha_\ell \rightarrow 0$ and hence $h_\ell \rightarrow k_\ell$. Inserting this form for $\hat{g}(\bm{r},\bm{r}^\prime ,z)$ into the radiation boundary condition in Eq.~(\ref{eq:psi}), we find that
\begin{align*}
{\Gamma}_{\ell m} &=  \frac{1}{\nu R}\frac{1}{h_{\ell}(\nu R)h_m(\nu R)}
\frac{2}{2\ell +1} \frac{2}{2m+1} (\bm{M}^{-1})_{\ell m} \\
&  -\frac{2}{2m+1} \frac{i_{\ell}(\nu R)}{h_{\ell}(\nu R)} \delta_{\ell m}
\end{align*}
where the $z$-dependent matrix $\bm{M}$ is defined as
\begin{align}
{M}_{\ell m} &= \frac{2 Q_{\ell}(\nu R)}{2 \ell +1} \delta_{\ell m} + \frac{k^0}{k_D} \int_0^1 d\mu \; P_m(\mu) P_{\ell}(\mu) \label{MatrixDef}
\end{align}
with
\begin{eqnarray*}
Q_{\ell}(\nu R) = \frac{\nu R \, \big( k_{\ell+1} (\nu R) + \alpha_l i_{\ell +1}(\nu R) \big)}
{k_{\ell}(\nu R) - \alpha_l i_{\ell}(\nu R)} - \ell.
\end{eqnarray*}
The matrix $\bm{M}$ defined above may be easily evaluated using the Wigner $3j$-symbols\cite{Messiah},
\begin{align}
\int_0^1 d\mu \, P_l(\mu) P_m(\mu) &= \sum_{n=|l-m|}^{|l+m|}
\begin{pmatrix}
2l & 2m & 2n \\
0 & 0 & 0
\end{pmatrix}^2 (2n+1) E_n
\label{eq:Mmatrix}
\end{align}
where $E_0 = 1$ and for $n \geq 1$,
\begin{align}
E_n &= \int_0^1 d\mu \; P_n(\mu) =  \frac{P_{n-1}(0) - P_{n+1}(0)}{2n+1}.
\label{eq:Evector}
\end{align}

With this form of the matrix $\boldsymbol{\Gamma}$, the Green function can be written as
\begin{eqnarray}
\hat{g}(\bm{r},\bm{r}^\prime ,z) &=& \frac{\nu}{2\pi} \sum_{\ell = 0}^\infty \frac{2\ell + 1}{2}
P_\ell (\mu) \hat{g}_{\ell}^d(r,r^\prime) P_{\ell}(\mu^\prime) \label{GreensFunction} \\
&&+ \frac{1}{2\pi R} \sum_{\ell, m} P_{\ell}(\mu) \frac{h_{\ell}(\nu r)}{h_{\ell}(\nu R)}
(\bm{M}^{-1})_{\ell m} \frac{h_m(\nu r^\prime)}{h_m(\nu R)} P_m(\mu^\prime) \nonumber
\end{eqnarray}
where
\begin{eqnarray}
&&\hat{g}_{\ell}^d (r,r^\prime) = i_{\ell}(\nu r) h_{\ell}(\nu r^\prime) H(r^\prime - r) \\
&&\quad+h_{\ell}(\nu r) i_{\ell}(\nu r^\prime) H(r-r^\prime) - h_{\ell}(\nu r)
\frac{i_{\ell}(\nu R)}{h_{\ell}(\nu R)} h_{\ell}(\nu r^\prime).\nonumber
\end{eqnarray}

Using the Green function for a spatially uniform initial fluctuation $\psi(0) = \delta N_A(0)/V$, in the limit $R_m\rightarrow \infty$ we obtain the $z$-dependent concentration fluctuation fields and the particle number fluctuations from
\begin{align}
\widehat{\psi}(\bm{r}, z)
&= \frac{\psi(0)}{D\nu^2} \bigg[
1 - \frac{k_{0}(\nu r)}{k_0 (\nu R)} \label{eq:ca} \\
& \qquad + 2 (1+\nu R) \sum_{\ell = 0}^\infty
 \frac{k_{\ell}(\nu r)}{k_{\ell}(\nu R)} (\bm{M}^{-1})_{0 \ell} P_{\ell}(\mu) \bigg].
\nonumber
\end{align}
The radiation boundary condition implies that
\begin{align*}
1 - 2(1+\nu R) \bm{M}^{-1}_{00} = \frac{k^0}{k_D} \sum_{\ell = 0}^\infty (\bm{M}^{-1})_{0 \ell} E_\ell
= \frac{k^0}{k_D} a_{0}(z),
\end{align*}
where we have defined the vector components $a_{k}(z) = \sum_{\ell} (\bm{M}^{-1})_{k \ell}E_{\ell}$. From the integral of Eq.~(\ref{eq:ca}) over the volume outside the Janus particle we may obtain Eq.~(\ref{Na_z}) of the main text and, setting $k_{\pm 2}=0$, Eq.~(\ref{Na_z-nb}).


\begin{thebibliography}{32}%
\makeatletter
\providecommand \@ifxundefined [1]{%
 \@ifx{#1\undefined}
}%
\providecommand \@ifnum [1]{%
 \ifnum #1\expandafter \@firstoftwo
 \else \expandafter \@secondoftwo
 \fi
}%
\providecommand \@ifx [1]{%
 \ifx #1\expandafter \@firstoftwo
 \else \expandafter \@secondoftwo
 \fi
}%
\providecommand \natexlab [1]{#1}%
\providecommand \enquote  [1]{``#1''}%
\providecommand \bibnamefont  [1]{#1}%
\providecommand \bibfnamefont [1]{#1}%
\providecommand \citenamefont [1]{#1}%
\providecommand \href@noop [0]{\@secondoftwo}%
\providecommand \href [0]{\begingroup \@sanitize@url \@href}%
\providecommand \@href[1]{\@@startlink{#1}\@@href}%
\providecommand \@@href[1]{\endgroup#1\@@endlink}%
\providecommand \@sanitize@url [0]{\catcode `\\12\catcode `\$12\catcode
  `\&12\catcode `\#12\catcode `\^12\catcode `\_12\catcode `\%12\relax}%
\providecommand \@@startlink[1]{}%
\providecommand \@@endlink[0]{}%
\providecommand \url  [0]{\begingroup\@sanitize@url \@url }%
\providecommand \@url [1]{\endgroup\@href {#1}{\urlprefix }}%
\providecommand \urlprefix  [0]{URL }%
\providecommand \Eprint [0]{\href }%
\providecommand \doibase [0]{http://dx.doi.org/}%
\providecommand \selectlanguage [0]{\@gobble}%
\providecommand \bibinfo  [0]{\@secondoftwo}%
\providecommand \bibfield  [0]{\@secondoftwo}%
\providecommand \translation [1]{[#1]}%
\providecommand \BibitemOpen [0]{}%
\providecommand \bibitemStop [0]{}%
\providecommand \bibitemNoStop [0]{.\EOS\space}%
\providecommand \EOS [0]{\spacefactor3000\relax}%
\providecommand \BibitemShut  [1]{\csname bibitem#1\endcsname}%
\let\auto@bib@innerbib\@empty
\bibitem [{\citenamefont {Alberts}\ \emph {et~al.}(2002)\citenamefont
  {Alberts}, \citenamefont {Bray}, \citenamefont {Lewis}, \citenamefont {Raff},
  \citenamefont {Roberts},\ and\ \citenamefont {Watson}}]{alberts-cell}%
  \BibitemOpen
  \bibfield  {author} {\bibinfo {author} {\bibfnamefont {B.}~\bibnamefont
  {Alberts}}, \bibinfo {author} {\bibfnamefont {D.}~\bibnamefont {Bray}},
  \bibinfo {author} {\bibfnamefont {J.}~\bibnamefont {Lewis}}, \bibinfo
  {author} {\bibfnamefont {M.}~\bibnamefont {Raff}}, \bibinfo {author}
  {\bibfnamefont {K.}~\bibnamefont {Roberts}}, \ and\ \bibinfo {author}
  {\bibfnamefont {J.~D.}\ \bibnamefont {Watson}},\ }\href@noop {} {\emph
  {\bibinfo {title} {Molecular Biology of the Cell}}},\ \bibinfo {edition}
  {3rd}\ ed.\ (\bibinfo  {publisher} {Garland Science},\ \bibinfo {year}
  {2002})\BibitemShut {NoStop}%
\bibitem [{\citenamefont {Jones}(2004)}]{Jones-book}%
  \BibitemOpen
  \bibfield  {author} {\bibinfo {author} {\bibfnamefont {R.~A.~L.}\
  \bibnamefont {Jones}},\ }\href@noop {} {\emph {\bibinfo {title} {Soft
  Machines: Nanotechnology and Life}}}\ (\bibinfo  {publisher} {Oxford
  University Press},\ \bibinfo {address} {Oxford},\ \bibinfo {year}
  {2004})\BibitemShut {NoStop}%
\bibitem [{\citenamefont {Berg}(1975)}]{berg75}%
  \BibitemOpen
  \bibfield  {author} {\bibinfo {author} {\bibfnamefont {H.~C.}\ \bibnamefont
  {Berg}},\ }\href {\doibase 10.1146/annurev.bb.04.060175.001003} {\bibfield
  {journal} {\bibinfo  {journal} {Annu. Rev. Biophys. and Bioeng.}\ }\textbf
  {\bibinfo {volume} {4}},\ \bibinfo {pages} {119} (\bibinfo {year}
  {1975})}\BibitemShut {NoStop}%
\bibitem [{\citenamefont {Berg}(2004)}]{berg04}%
  \BibitemOpen
  \bibfield  {author} {\bibinfo {author} {\bibfnamefont {H.~C.}\ \bibnamefont
  {Berg}},\ }\href@noop {} {\emph {\bibinfo {title} {E. coli in Motion}}}\
  (\bibinfo  {publisher} {Springer},\ \bibinfo {address} {New York},\ \bibinfo
  {year} {2004})\BibitemShut {NoStop}%
\bibitem [{\citenamefont {Kay}, \citenamefont {Leigh},\ and\ \citenamefont
  {Zerbetto}(2007)}]{kay:07}%
  \BibitemOpen
  \bibfield  {author} {\bibinfo {author} {\bibfnamefont {E.~R.}\ \bibnamefont
  {Kay}}, \bibinfo {author} {\bibfnamefont {D.~A.}\ \bibnamefont {Leigh}}, \
  and\ \bibinfo {author} {\bibfnamefont {F.}~\bibnamefont {Zerbetto}},\
  }\href@noop {} {\bibfield  {journal} {\bibinfo  {journal} {Angew. Chem. Int.
  Ed.}\ }\textbf {\bibinfo {volume} {46}},\ \bibinfo {pages} {72} (\bibinfo
  {year} {2007})}\BibitemShut {NoStop}%
\bibitem [{\citenamefont {Wang}(2013)}]{wangbook:13}%
  \BibitemOpen
  \bibfield  {author} {\bibinfo {author} {\bibfnamefont {J.}~\bibnamefont
  {Wang}},\ }\href@noop {} {\emph {\bibinfo {title} {Nanomachines: Fundamentals
  and Applications}}}\ (\bibinfo  {publisher} {Wiley-VCH},\ \bibinfo {address}
  {Weinheim, Germany},\ \bibinfo {year} {2013})\BibitemShut {NoStop}%
\bibitem [{\citenamefont {Wang}\ \emph {et~al.}(2013)\citenamefont {Wang},
  \citenamefont {Duan}, \citenamefont {Ahmed}, \citenamefont {Mallouk},\ and\
  \citenamefont {Sen}}]{wang:13}%
  \BibitemOpen
  \bibfield  {author} {\bibinfo {author} {\bibfnamefont {W.}~\bibnamefont
  {Wang}}, \bibinfo {author} {\bibfnamefont {W.}~\bibnamefont {Duan}}, \bibinfo
  {author} {\bibfnamefont {S.}~\bibnamefont {Ahmed}}, \bibinfo {author}
  {\bibfnamefont {T.~E.}\ \bibnamefont {Mallouk}}, \ and\ \bibinfo {author}
  {\bibfnamefont {A.}~\bibnamefont {Sen}},\ }\href@noop {} {\bibfield
  {journal} {\bibinfo  {journal} {Nano Today}\ }\textbf {\bibinfo {volume}
  {8}},\ \bibinfo {pages} {531} (\bibinfo {year} {2013})}\BibitemShut {NoStop}%
\bibitem [{\citenamefont {S\'{a}nchez}, \citenamefont {Soler},\ and\
  \citenamefont {Katuri}(2014)}]{sanchez:14}%
  \BibitemOpen
  \bibfield  {author} {\bibinfo {author} {\bibfnamefont {S.}~\bibnamefont
  {S\'{a}nchez}}, \bibinfo {author} {\bibfnamefont {L.}~\bibnamefont {Soler}},
  \ and\ \bibinfo {author} {\bibfnamefont {J.}~\bibnamefont {Katuri}},\
  }\href@noop {} {\bibfield  {journal} {\bibinfo  {journal} {Angew. Chem. Int.
  Ed.}\ }\textbf {\bibinfo {volume} {53}},\ \bibinfo {pages} {2} (\bibinfo
  {year} {2014})}\BibitemShut {NoStop}%
\bibitem [{\citenamefont {Derjaguin}\ \emph {et~al.}(1947)\citenamefont
  {Derjaguin}, \citenamefont {Sidorenkov}, \citenamefont {Zubashchenkov},\ and\
  \citenamefont {Kiseleva}}]{derjaguin:47}%
  \BibitemOpen
  \bibfield  {author} {\bibinfo {author} {\bibfnamefont {B.~V.}\ \bibnamefont
  {Derjaguin}}, \bibinfo {author} {\bibfnamefont {G.~P.}\ \bibnamefont
  {Sidorenkov}}, \bibinfo {author} {\bibfnamefont {E.~A.}\ \bibnamefont
  {Zubashchenkov}}, \ and\ \bibinfo {author} {\bibfnamefont {E.~V.}\
  \bibnamefont {Kiseleva}},\ }\href@noop {} {\bibfield  {journal} {\bibinfo
  {journal} {Kolloidn. Zh.}\ }\textbf {\bibinfo {volume} {9}},\ \bibinfo
  {pages} {335} (\bibinfo {year} {1947})}\BibitemShut {NoStop}%
\bibitem [{\citenamefont {Dukhin}\ and\ \citenamefont
  {Derjaguin}(1974)}]{derjaguin:74}%
  \BibitemOpen
  \bibfield  {author} {\bibinfo {author} {\bibfnamefont {S.~S.}\ \bibnamefont
  {Dukhin}}\ and\ \bibinfo {author} {\bibfnamefont {B.~V.}\ \bibnamefont
  {Derjaguin}},\ }\href@noop {} {\emph {\bibinfo {title} {in Surface and
  Colloid Sicence, ed. E. Matijevic}}},\ Vol.~\bibinfo {volume} {7}\ (\bibinfo
  {publisher} {Wiley},\ \bibinfo {year} {1974})\ p.\ \bibinfo {pages}
  {365}\BibitemShut {NoStop}%
\bibitem [{\citenamefont {Anderson}, \citenamefont {Lowell},\ and\
  \citenamefont {Prieve}(1982)}]{ALP82}%
  \BibitemOpen
  \bibfield  {author} {\bibinfo {author} {\bibfnamefont {J.~L.}\ \bibnamefont
  {Anderson}}, \bibinfo {author} {\bibfnamefont {M.~E.}\ \bibnamefont
  {Lowell}}, \ and\ \bibinfo {author} {\bibfnamefont {D.~C.}\ \bibnamefont
  {Prieve}},\ }\href@noop {} {\bibfield  {journal} {\bibinfo  {journal} {J.
  Fluid Mech.}\ }\textbf {\bibinfo {volume} {117}},\ \bibinfo {pages} {107}
  (\bibinfo {year} {1982})}\BibitemShut {NoStop}%
\bibitem [{\citenamefont {Anderson}(1989)}]{A89}%
  \BibitemOpen
  \bibfield  {author} {\bibinfo {author} {\bibfnamefont {J.~L.}\ \bibnamefont
  {Anderson}},\ }\href@noop {} {\bibfield  {journal} {\bibinfo  {journal} {Ann.
  Rev. Fluid Mech.}\ }\textbf {\bibinfo {volume} {21}},\ \bibinfo {pages} {61}
  (\bibinfo {year} {1989})}\BibitemShut {NoStop}%
\bibitem [{\citenamefont {Anderson}\ and\ \citenamefont {Prieve}(1991)}]{AP91}%
  \BibitemOpen
  \bibfield  {author} {\bibinfo {author} {\bibfnamefont {J.~L.}\ \bibnamefont
  {Anderson}}\ and\ \bibinfo {author} {\bibfnamefont {D.~C.}\ \bibnamefont
  {Prieve}},\ }\href@noop {} {\bibfield  {journal} {\bibinfo  {journal}
  {Langmuir}\ }\textbf {\bibinfo {volume} {7}},\ \bibinfo {pages} {403}
  (\bibinfo {year} {1991})}\BibitemShut {NoStop}%
\bibitem [{\citenamefont {Golestanian}, \citenamefont {Liverpool},\ and\
  \citenamefont {Ajdari}(2005)}]{Golestanian_etal_05}%
  \BibitemOpen
  \bibfield  {author} {\bibinfo {author} {\bibfnamefont {R.}~\bibnamefont
  {Golestanian}}, \bibinfo {author} {\bibfnamefont {T.~B.}\ \bibnamefont
  {Liverpool}}, \ and\ \bibinfo {author} {\bibfnamefont {A.}~\bibnamefont
  {Ajdari}},\ }\href@noop {} {\bibfield  {journal} {\bibinfo  {journal} {Phys.
  Rev. Lett.}\ }\textbf {\bibinfo {volume} {94}},\ \bibinfo {pages} {220801}
  (\bibinfo {year} {2005})}\BibitemShut {NoStop}%
\bibitem [{\citenamefont {Kapral}(2013)}]{kapral:13}%
  \BibitemOpen
  \bibfield  {author} {\bibinfo {author} {\bibfnamefont {R.}~\bibnamefont
  {Kapral}},\ }\href@noop {} {\bibfield  {journal} {\bibinfo  {journal} {J.
  Chem. Phys.}\ }\textbf {\bibinfo {volume} {138}},\ \bibinfo {pages} {020901}
  (\bibinfo {year} {2013})}\BibitemShut {NoStop}%
\bibitem [{\citenamefont {Gaspard}\ and\ \citenamefont {Kapral}(2017)}]{GK17}%
  \BibitemOpen
  \bibfield  {author} {\bibinfo {author} {\bibfnamefont {P.}~\bibnamefont
  {Gaspard}}\ and\ \bibinfo {author} {\bibfnamefont {R.}~\bibnamefont
  {Kapral}},\ }\href@noop {} {\bibfield  {journal} {\bibinfo  {journal} {J.
  Chem. Phys.}\ }\textbf {\bibinfo {volume} {147}},\ \bibinfo {pages} {211101}
  (\bibinfo {year} {2017})}\BibitemShut {NoStop}%
\bibitem [{\citenamefont {Gaspard}\ and\ \citenamefont {Kapral}(2018)}]{GK18a}%
  \BibitemOpen
  \bibfield  {author} {\bibinfo {author} {\bibfnamefont {P.}~\bibnamefont
  {Gaspard}}\ and\ \bibinfo {author} {\bibfnamefont {R.}~\bibnamefont
  {Kapral}},\ }\href@noop {} {\bibfield  {journal} {\bibinfo  {journal}
  {arXiv:1801.00766}\ } (\bibinfo {year} {2018})}\BibitemShut {NoStop}%
\bibitem [{\citenamefont {de~Buyl}\ and\ \citenamefont
  {Kapral}(2013)}]{debuyl:13}%
  \BibitemOpen
  \bibfield  {author} {\bibinfo {author} {\bibfnamefont {P.}~\bibnamefont
  {de~Buyl}}\ and\ \bibinfo {author} {\bibfnamefont {R.}~\bibnamefont
  {Kapral}},\ }\href@noop {} {\bibfield  {journal} {\bibinfo  {journal}
  {Nanoscale}\ }\textbf {\bibinfo {volume} {5}},\ \bibinfo {pages} {1337}
  (\bibinfo {year} {2013})}\BibitemShut {NoStop}%
\bibitem [{\citenamefont {de~Buyl}(2018)}]{debuyl:18}%
  \BibitemOpen
  \bibfield  {author} {\bibinfo {author} {\bibfnamefont {P.}~\bibnamefont
  {de~Buyl}},\ }\href@noop {} {\bibfield  {journal} {\bibinfo  {journal}
  {arXiv:1802.03264v1}\ } (\bibinfo {year} {2018})}\BibitemShut {NoStop}%
\bibitem [{\citenamefont {Malevanets}\ and\ \citenamefont
  {Kapral}(1999)}]{Malevanets_Kapral_99}%
  \BibitemOpen
  \bibfield  {author} {\bibinfo {author} {\bibfnamefont {A.}~\bibnamefont
  {Malevanets}}\ and\ \bibinfo {author} {\bibfnamefont {R.}~\bibnamefont
  {Kapral}},\ }\href@noop {} {\bibfield  {journal} {\bibinfo  {journal} {J.
  Chem. Phys.}\ }\textbf {\bibinfo {volume} {110}},\ \bibinfo {pages} {8605}
  (\bibinfo {year} {1999})}\BibitemShut {NoStop}%
\bibitem [{\citenamefont {Malevanets}\ and\ \citenamefont
  {Kapral}(2000)}]{Malevanets_Kapral_00}%
  \BibitemOpen
  \bibfield  {author} {\bibinfo {author} {\bibfnamefont {A.}~\bibnamefont
  {Malevanets}}\ and\ \bibinfo {author} {\bibfnamefont {R.}~\bibnamefont
  {Kapral}},\ }\href@noop {} {\bibfield  {journal} {\bibinfo  {journal} {J.
  Chem. Phys.}\ }\textbf {\bibinfo {volume} {112}},\ \bibinfo {pages} {7260}
  (\bibinfo {year} {2000})}\BibitemShut {NoStop}%
\bibitem [{\citenamefont {Kapral}(2008)}]{Kapral_08}%
  \BibitemOpen
  \bibfield  {author} {\bibinfo {author} {\bibfnamefont {R.}~\bibnamefont
  {Kapral}},\ }\href@noop {} {\bibfield  {journal} {\bibinfo  {journal} {Adv.
  Chem. Phys.}\ }\textbf {\bibinfo {volume} {140}},\ \bibinfo {pages} {89}
  (\bibinfo {year} {2008})}\BibitemShut {NoStop}%
\bibitem [{\citenamefont {Zwanzig}(2001)}]{zwanzigbook:01}%
  \BibitemOpen
  \bibfield  {author} {\bibinfo {author} {\bibfnamefont {R.}~\bibnamefont
  {Zwanzig}},\ }\href@noop {} {\emph {\bibinfo {title} {Nonequilibrium
  Statistical Mechanics}}}\ (\bibinfo  {publisher} {Oxford University Press},\
  \bibinfo {address} {New York},\ \bibinfo {year} {2001})\BibitemShut {NoStop}%
\bibitem [{\citenamefont {Kapral}(1981)}]{kapral:81}%
  \BibitemOpen
  \bibfield  {author} {\bibinfo {author} {\bibfnamefont {R.}~\bibnamefont
  {Kapral}},\ }\href@noop {} {\bibfield  {journal} {\bibinfo  {journal} {Adv.
  Chem. Phys.}\ }\textbf {\bibinfo {volume} {48}},\ \bibinfo {pages} {71}
  (\bibinfo {year} {1981})}\BibitemShut {NoStop}%
\bibitem [{\citenamefont {Onsager}(1931{\natexlab{a}})}]{Onsager:31a}%
  \BibitemOpen
  \bibfield  {author} {\bibinfo {author} {\bibfnamefont {L.}~\bibnamefont
  {Onsager}},\ }\href@noop {} {\bibfield  {journal} {\bibinfo  {journal} {Phys.
  Rev.}\ }\textbf {\bibinfo {volume} {37}},\ \bibinfo {pages} {405} (\bibinfo
  {year} {1931}{\natexlab{a}})}\BibitemShut {NoStop}%
\bibitem [{\citenamefont {Onsager}(1931{\natexlab{b}})}]{Onsager:31b}%
  \BibitemOpen
  \bibfield  {author} {\bibinfo {author} {\bibfnamefont {L.}~\bibnamefont
  {Onsager}},\ }\href@noop {} {\bibfield  {journal} {\bibinfo  {journal} {Phys.
  Rev.}\ }\textbf {\bibinfo {volume} {38}},\ \bibinfo {pages} {2265} (\bibinfo
  {year} {1931}{\natexlab{b}})}\BibitemShut {NoStop}%
\bibitem [{\citenamefont {Collins}\ and\ \citenamefont
  {Kimball}(1949)}]{Collins_Kimball_49}%
  \BibitemOpen
  \bibfield  {author} {\bibinfo {author} {\bibfnamefont {F.~C.}\ \bibnamefont
  {Collins}}\ and\ \bibinfo {author} {\bibfnamefont {G.~E.}\ \bibnamefont
  {Kimball}},\ }\href@noop {} {\bibfield  {journal} {\bibinfo  {journal} {J.
  Colloid Sci.}\ }\textbf {\bibinfo {volume} {4}},\ \bibinfo {pages} {425}
  (\bibinfo {year} {1949})}\BibitemShut {NoStop}%
\bibitem [{\citenamefont {Ajdari}\ and\ \citenamefont {Bocquet}(2006)}]{AB06}%
  \BibitemOpen
  \bibfield  {author} {\bibinfo {author} {\bibfnamefont {A.}~\bibnamefont
  {Ajdari}}\ and\ \bibinfo {author} {\bibfnamefont {L.}~\bibnamefont
  {Bocquet}},\ }\href@noop {} {\bibfield  {journal} {\bibinfo  {journal} {Phys.
  Rev. Lett.}\ }\textbf {\bibinfo {volume} {96}},\ \bibinfo {pages} {186102}
  (\bibinfo {year} {2006})}\BibitemShut {NoStop}%
\bibitem [{\citenamefont {Romanczuk}\ \emph {et~al.}(2012)\citenamefont
  {Romanczuk}, \citenamefont {B\"ar}, \citenamefont {Ebeling}, \citenamefont
  {Lindner},\ and\ \citenamefont {Schimansky-Geier}}]{RBELS12}%
  \BibitemOpen
  \bibfield  {author} {\bibinfo {author} {\bibfnamefont {P.}~\bibnamefont
  {Romanczuk}}, \bibinfo {author} {\bibfnamefont {M.}~\bibnamefont {B\"ar}},
  \bibinfo {author} {\bibfnamefont {W.}~\bibnamefont {Ebeling}}, \bibinfo
  {author} {\bibfnamefont {B.}~\bibnamefont {Lindner}}, \ and\ \bibinfo
  {author} {\bibfnamefont {L.}~\bibnamefont {Schimansky-Geier}},\ }\href@noop
  {} {\bibfield  {journal} {\bibinfo  {journal} {Eur. Phys. J. Special Topics}\
  }\textbf {\bibinfo {volume} {202}},\ \bibinfo {pages} {1} (\bibinfo {year}
  {2012})}\BibitemShut {NoStop}%
\bibitem [{\citenamefont {Rohlf}, \citenamefont {Fraser},\ and\ \citenamefont
  {Kapral}(2008)}]{Rohlf_etal_08}%
  \BibitemOpen
  \bibfield  {author} {\bibinfo {author} {\bibfnamefont {K.}~\bibnamefont
  {Rohlf}}, \bibinfo {author} {\bibfnamefont {S.}~\bibnamefont {Fraser}}, \
  and\ \bibinfo {author} {\bibfnamefont {R.}~\bibnamefont {Kapral}},\
  }\href@noop {} {\bibfield  {journal} {\bibinfo  {journal} {Comput. Phys.
  Commun.}\ }\textbf {\bibinfo {volume} {179}},\ \bibinfo {pages} {132}
  (\bibinfo {year} {2008})}\BibitemShut {NoStop}%
\bibitem [{\citenamefont {Gompper}\ \emph {et~al.}(2009)\citenamefont
  {Gompper}, \citenamefont {Ihle}, \citenamefont {Kroll},\ and\ \citenamefont
  {Winkler}}]{gompper:09}%
  \BibitemOpen
  \bibfield  {author} {\bibinfo {author} {\bibfnamefont {G.}~\bibnamefont
  {Gompper}}, \bibinfo {author} {\bibfnamefont {T.}~\bibnamefont {Ihle}},
  \bibinfo {author} {\bibfnamefont {D.~M.}\ \bibnamefont {Kroll}}, \ and\
  \bibinfo {author} {\bibfnamefont {R.~G.}\ \bibnamefont {Winkler}},\
  }\href@noop {} {\bibfield  {journal} {\bibinfo  {journal} {Adv. Polym. Sci.}\
  }\textbf {\bibinfo {volume} {221}},\ \bibinfo {pages} {1} (\bibinfo {year}
  {2009})}\BibitemShut {NoStop}%
\bibitem [{\citenamefont {Messiah}(1962)}]{Messiah}%
  \BibitemOpen
  \bibfield  {author} {\bibinfo {author} {\bibfnamefont {A.}~\bibnamefont
  {Messiah}},\ }\href@noop {} {\emph {\bibinfo {title} {Quantum Mechanics, Vol.
  2}}}\ (\bibinfo  {publisher} {North Holland},\ \bibinfo {address} {Amsterdam,
  Netherlands},\ \bibinfo {year} {1962})\ pp.\ \bibinfo {pages}
  {1054--1060}\BibitemShut {NoStop}%
\end{thebibliography}

\providecommand{\noopsort}[1]{}\providecommand{\singleletter}[1]{#1}

\end{document}